\documentclass[11pt]{article}
\usepackage{amsfonts}
\usepackage{amssymb}
\usepackage{mathrsfs}
\usepackage{graphicx}
\usepackage{amsmath,amsthm,amssymb,amscd}
\usepackage[all]{xy}
\usepackage{hyperref}
\usepackage{multirow}
\usepackage{color}
\usepackage{float}
\usepackage{mathtools,slashed}
\usepackage{bbold}
\usepackage{bbm}
\usepackage{slashed}
\usepackage{cancel}
\usepackage[normalem]{ulem}
\usepackage{cite}

\usepackage{comment}

\newcommand\numberthis{\addtocounter{equation}{1}\tag{\theequation}} % For numberthis in align

\usepackage{subcaption} % For subfigures
\usepackage[T1]{fontenc} % For latin characters
\usepackage[toc,page]{appendix} % For appendices

\usepackage[nottoc,numbib]{tocbibind}

\allowdisplaybreaks % Page breaking for equations!

\usepackage{adjustbox}
\usepackage{amsxtra,graphics,epsfig,bm,tikz,xfrac,lscape}
\usetikzlibrary{decorations.pathmorphing}
\usetikzlibrary{decorations.markings}
\usetikzlibrary{arrows, decorations.markings, calc, fadings, decorations.pathreplacing, patterns, decorations.pathmorphing, positioning}

\usetikzlibrary{positioning,shapes}
\usetikzlibrary{chains}
\usetikzlibrary{arrows,fit,decorations.pathreplacing}
\tikzstyle{every picture}+=[remember picture]
\tikzstyle{na} = [baseline=-.5ex]

\renewcommand{\theequation}{\thesection.\arabic{equation}}

\newcommand{\nn}{\nonumber}

\def\eqa{\begin{eqnarray}}
\def\eqae{\end{eqnarray}}
\def\eq{\begin{equation}}
\def\eqe{\end{equation}}
\def\be{\begin{equation}}
\def\ee{\end{equation}}
\def\bea{\begin{eqnarray}}
\def\eea{\end{eqnarray}}
\def\ba{\begin{array}}
\def\ea{\end{array}}
\def\bd{\begin{displaymath}}
\def\ed{\end{displaymath}}

\def\>{\rangle}
\def\<{\langle}

\def\i{\iota}

\numberwithin{equation}{section}

\newcommand{\fft}[2]{\frac{#1}{#2}}
\newcommand{\ft}[2]{{\textstyle\frac{#1}{#2}}}

\definecolor{darkblue}{rgb}{0,0,0.5}
\definecolor{darkred}{rgb}{0.5,0,0}
\definecolor{darkgreen}{rgb}{0,0.5,0}
\definecolor{orange}{rgb}{0.9,0.58,0}

\begin{document}

\begin{titlepage}
\hfill LCTP-23-18

\vskip 1 cm

\begin{center}
%{\Large \bf  }\\

\vskip .7cm

{\Large \bf Discontinuity in RG Flows Across Dimensions:\\
\vskip .7 cm
Entanglement, Anomaly Coefficients and Geometry}\\

\vskip .7cm

\end{center}

\vskip .7 cm

\vskip 1 cm
\begin{center}
{\large  \bf  Jos\'e de-la-Cruz-Moreno$^a$, James T. Liu$^a$ and Leopoldo A. Pando Zayas$^{a,b}$}\\
\end{center}

\vskip .4cm \centerline{\it ${}^a$ Leinweber Center for Theoretical Physics, Randall Laboratory of Physics}
\centerline{ \it The University of Michigan, Ann Arbor, MI 48109-1040}

\bigskip\bigskip

\centerline{\it ${}^b$ The Abdus Salam International Centre for Theoretical Physics}
\centerline{\it  Strada Costiera 11,  34014 Trieste, Italy}

\bigskip\bigskip
\vskip 1.5 cm
\begin{abstract}
We study the entanglement entropy associated with a holographic RG flow from $\textrm{AdS}_7$ to $\textrm{AdS}_{4} \times \mathbb{H}_3$, where $\mathbb{H}_3$ is a $3$-dimensional hyperbolic manifold with curvature $\kappa$. The dual six-dimensional RG flow is disconnected from Lorentz-invariant flows. In this context we address various notions of central charges and identify a monotonic candidate $c$-function that captures IR aspects of the flow. The UV behavior of the holographic entanglement entropy and, in particular its universal term, display an interesting dependence on the curvature, $\kappa$. We then contrast our holographic results with existing field theory computations in six dimensions and find a series of new corrections  in curvature to the universal term in the entanglement entropy.
\end{abstract}

\vskip  1.5 cm
{\tt  josedcm@umich.edu, jimliu@umich.edu, lpandoz@umich.edu}
\end{titlepage}

\tableofcontents

%%%%%%%%%%%%%%%%%%%%%%%%%%%%%%%%%%%%%%%%%%%%%%%%%%%%%%%%%%%%%%%%%%%%%%%%%%%%%%%%%%%%%%%%%%%%%%%%%%%%%%%%%%%%%%%%%%%%%%%%%%%%%%%%%%%%%%%%%%%%%%%%%%%%%%%%%%%%%%%%%%%%%%%%%%%%%%%%%%%%%%%
\section{Introduction}
\label{introduction}
Quantum field theory provides a mathematical framework to investigate physical phenomena at different energy scales. Its organizing principle is the renormalization group (RG) flow  which describes the theory as a flow from a high energy scale, namely the ultraviolet (UV) regime, to a low energy scale, namely the infrared (IR) regime.  The central physics idea can be framed as coarse-graining the degrees of freedom \cite{Wilson:1973jj}. Precise statements regarding the decreasing of the effective number of  degrees of freedom as one approaches the IR are highly dependent on the spacetime dimension of the field theory in question. As a proxy for the number of degrees of freedom, one ideally searches for $c$-functions, {\it i.e.}, functions of the coupling constants which are monotonically decreasing along the RG flow, and that interpolate between UV and IR central charges. 

Monotonicity theorems have been established in various  dimensions utilizing different field-theoretic techniques. In  two dimensions the $c$-theorem was proven by Zamolodchikov exploiting properties of two-point functions of the stress-energy tensor \cite{Zamolodchikov:1986gt}. In three dimensions important explorations towards the $F$-theorem include \cite{Klebanov:2011gs,Jafferis:2011zi,Jafferis:2010un}. The $F$-theorem was proven following entropic considerations in  \cite{Casini:2012ei}. In four dimensions an $a$-theorem was anticipated in  \cite{Cardy:1988cwa} and tackled perturbatively in \cite{Osborn:1989td,Jack:1990eb}. The $a$-theorem was settled non-perturbatively in  \cite{Komargodski:2011vj} using properties of a certain two-to-two scattering amplitude. A unified entropic framework for proving these theorems in dimensions two, three and four was presented in  \cite{Casini:2017roe,Casini:2017vbe}. Evidence for monotonicity has been presented for five-dimensional theories in  \cite{Jafferis:2012iv,Chang:2017cdx,Fluder:2020pym} and for six-dimensional theories in  \cite{Elvang:2012st,Heckman:2015axa,Cordova:2015fha}.  Despite lots of evidence, complete proofs in dimensions five and six are still lacking. None of these impressive developments, however, extends to situations where Lorentz invariance is broken.

The AdS/CFT correspondence posits a mathematical equivalence between certain gravity theories defined in asymptotically  ($d+1$)-dimensional anti-de Sitter spacetime $\textrm{AdS}_{d+1}$, and $d$-dimensional conformal field theories $\textrm{CFT}_{d}$ defined on its conformal boundary \cite{Maldacena:1997re}. Within this correspondence, the radial coordinate of $\textrm{AdS}_{d+1}$ encodes information about the energy scale in the dual $\textrm{CFT}_{d}$. Early on in the development of AdS/CFT, it was understood that gravity solutions which interpolate between two asymptotically anti-de Sitter spacetimes can be interpreted as RG flow between the associated dual conformal field theories \cite{Girardello:1998pd,Freedman:1999gp}. Furthermore, the AdS/CFT correspondence provides an approach for the entropic  exploration of monotonicity theorems through the computation of the holographic entanglement entropy via the Ryu-Takayanagi  prescription \cite{Ryu:2006bv,Ryu:2006ef}.

A particular class of RG flows that is amenable to a holographic treatment pertains to the so-called RG flows across dimensions. These flows were introduced by Maldacena and N\'u\~nez in \cite{Maldacena:2000mw} who considered placing ${\cal N}=4$  SYM and the M5 brane theory on Riemann surfaces and discussed both sides of the correspondence. Subsequent works have extended and clarified various aspects, including independent computations of the central charges at the fixed points using field-theoretic anomaly polynomial techniques and holographic computations of central charges \cite{Benini:2013cda,Benini:2015bwz}. Holographically, we are going to focus on the fact that the corresponding gravity solutions interpolate between two asymptotically AdS spacetimes of different dimensions, AdS$_{D+1}$ and AdS$_{d+1}$  (see  \cite{Maldacena:2000mw,Acharya:2000mu,Gauntlett:2000ng,Gauntlett:2001qs,Gauntlett:2001jj,Benini:2013cda,Benini:2015bwz,Bobev:2017uzs} for a partial list of such solutions). One natural question to pose in this context regards the existence of a generalized notion of a $c$-function. Such an exploration was initiated in \cite{GonzalezLezcano:2022mcd} and further developed in \cite{Deddo:2022wxj,Deddo:2023pid}.  Let us briefly recall some of the main results of \cite{GonzalezLezcano:2022mcd}  as those studies directly connect to the work pursued in this manuscript. There were three main examples of RG flows across dimensions discussed in \cite{GonzalezLezcano:2022mcd}. The first one concerns holographic RG flows from $\textrm{AdS}_5$ to $\textrm{AdS}_3 \times \Sigma_g$ associated with twisted partial compactifications of $4d$ $\mathcal{N}=4$ super Yang-Mills theory on a genus $g$ Riemann surface $\Sigma_g$, the second one refers to holographic RG flows from $\textrm{AdS}_7$ to $\textrm{AdS}_3 \times \mathbb{H}_4$ associated with partial compactifications of $6d$ $\mathcal{N}=(2,0)$ theories on a $4$-dimensional hyperbolic manifold $\mathbb{H}_4$, and the third one corresponds to holographic RG flows from $\textrm{AdS}_7$ to $\textrm{AdS}_5 \times T^2$ associated with twisted partial compactifications of $6d$ $\mathcal{N}=(2,0)$ theories on a 2-torus $T^2$. In the present paper we are interested in the construction of the holographic entanglement entropy associated with holographic RG flows from $\textrm{AdS}_7$ to $\textrm{AdS}_{4} \times \mathbb{H}_3$, where $\mathbb{H}_3$ is a $3$-dimensional hyperbolic manifold, the construction of candidate $c$-functions and general lessons for the entropy of 6d theories with explicitly broken Lorentz symmetry.

There are clear obstacles to the existence of a $c$-function across dimensions; they can be clearly identified in the case of compatifications with fluxes. For example, compactifying ${\cal N}=4$ SYM on $\Sigma_g$ with fluxes leads to a 2d central charge proportional to the genus of the Riemann surface \cite{Benini:2013cda}. As a result, this 2d central charge can be made arbitrarily large, even larger than the 4d central charge of the original theory.  More generally, it seems to be the case that compactifications with fluxes lead to a {\it new type of RG flow} that can be studied holographically but does not cleanly fall into the standard field-theoretic paradigm. The fact that the fluxes cannot be turned off in these compactifications points to certain discontinuities in the space of deformations. Namely, the field theoretic paradigm of deforming a conformal fixed point  by a relevant deformation that can be turned off does not apply. In this manuscript we explore a flow that is not {\it a priori} of this type as it is triggered by a scalar operator. We also wanted to explore a flow between even and odd dimensions, namely from a 6d field theory to a 3d theory, where reading the central charges has added challenges.

The rest of this paper is structured as follows. In section \ref{HolographicEE} we briefly review general aspects of RG flows across dimensions and their entropic description in holography.  In section \ref{HolographicEE7to4} we construct the holographic RG flows from $\textrm{AdS}_7$ to $\textrm{AdS}_{4} \times \mathbb{H}_3$ and present a combination of analytical and numerical results to establish the solution. Section \ref{HEE} is devoted to the construction of the holographic entanglement entropy. Section \ref{Thecfunction} focuses on the construction of the associated $c$-function. In section \ref{sec:anomalycoefs} we discuss  general aspects of anomalies for 6d field theories and broaden/generalize previous considerations in the literature \cite{Ryu:2006ef,Solodukhin:2008dh,Hung:2011xb}. Finally, in section \ref{conclusions} we present our conclusions and some venues for future work. We relegate a number of more technical aspects to a series of appendices.

%%%%%%%%%%%%%%%%%%%%%%%%%%%%%%%%%%%%%%%%%%%%%%%%%%%%%%%%%%%%%%%%%%%%%%%%%%%%%%%%%%%%%%%%%%%%%%%%%%%%%%%%%%%%%%%%%%%%%%%%%%%%%%%%%%%%%%%%%%%%%%%%%%%%%%%%%%%%%%%%%%%%%%%%%%%%%%%%%%%%%%%
\section{Discontinuity in RG flows across dimensions by fluxes and curvature}
\label{HolographicEE}

%%%%%%%%%%%%%%%%%%%%%%%%%%%%%%%%%%%%%%%%%%%%
%\subsection{Properties of flows across dimensions}

Let us highlight some of the properties of RG flows across dimensions that distinguish them from RG flows in the same spacetime dimension, that is, from Lorentz-invariant RG flows. Some of the properties are made explicit in the holographic context but it is worth stating them in a more quantum field theoretic framework. 

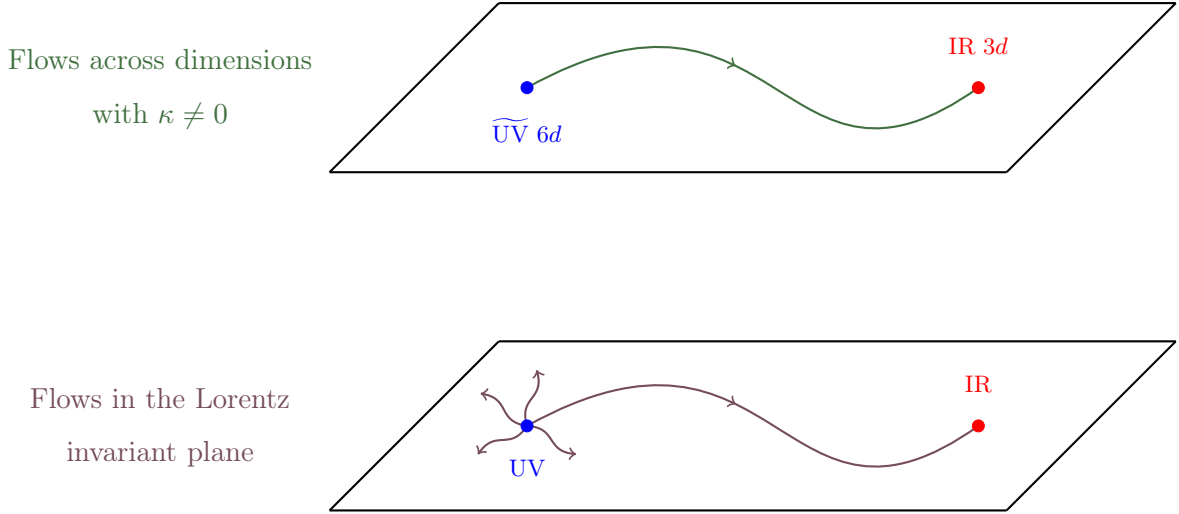
\begin{figure}[ht]
    \begin{adjustbox}{minipage=\textwidth, scale=1}
    \centering
    \begin{tikzpicture}[scale=0.75]
        %\draw[step=1cm, gray!25!, thick] (-6,-1) grid (16,10);
	    	
	    % Plane 1
		%\draw[thick, black] (3,10) -- (15,10);
		%\draw[thick, black] (0,7) -- (12,7);
		%\draw[thick, black] (3,10) -- (0,7);
		%\draw[thick, black] (15,10) -- (12,7);
		
		%\begin{scope}[decoration={markings,mark=at position 0.45 with {\arrow{>}}}]
		%\draw[thick, green!25!black!75, postaction={decorate}] (3.5,8.5) .. controls (8,11) and (8,6) .. (11.5,8.5);
		%\end{scope}

		%\filldraw[blue] (3.5,8.5) circle (3pt);
		%\filldraw[red] (11.5,8.5) circle (3pt);
		
		%\node[label=:] (A1) at (-3,9) {\textcolor{green!25!black!75}{Flows across dimensions}};
	    %\node[label=:] (B1) at (-3,8) {\textcolor{green!25!black!75}{with $\kappa \neq 0$}};
	    %\node[label=:] (C1) at (3.5,7.75) {\footnotesize \textcolor{blue}{UV $6d$}};
	    %\node[label=:] (D1) at (11.5,9.25) {\footnotesize \textcolor{red}{IR $3d$}};
	    
		% Plane 1
		\begin{scope}[shift={(0cm,-1cm)}]
		
		\draw[thick, black] (3,10) -- (15,10);
		\draw[thick, black] (0,7) -- (12,7);
		\draw[thick, black] (3,10) -- (0,7);
		\draw[thick, black] (15,10) -- (12,7);
		
		\begin{scope}[decoration={markings,mark=at position 0.45 with {\arrow{>}}}]
		\draw[thick, green!25!black!75, postaction={decorate}] (3.5,8.5) .. controls (8,11) and (8,6) .. (11.5,8.5);
		\end{scope}

		\filldraw[blue] (3.5,8.5) circle (3pt);
		\filldraw[red] (11.5,8.5) circle (3pt);
		
		\node[label=:] (A1) at (-3,9) {\textcolor{green!25!black!75}{Flows across dimensions}};
	    \node[label=:] (B1) at (-3,8) {\textcolor{green!25!black!75}{with $\kappa \neq 0$}};
	    \node[label=:] (C1) at (3.5,7.75) {\footnotesize \textcolor{blue}{$\widetilde{\rm UV} \,\, 6d$}};
	    \node[label=:] (D1) at (11.5,9.25) {\footnotesize \textcolor{red}{IR $3d$}};		
		
		\end{scope}
	    
	    % Plane 2
	    \draw[thick, black] (3,3) -- (15,3);
	    \draw[thick, black] (0,0) -- (12,0);
	    \draw[thick, black] (3,3) -- (0,0);
	    \draw[thick, black] (15,3) -- (12,0);
	    
		\begin{scope}[decoration={markings,mark=at position 0.45 with {\arrow{>}}}, shift={(0cm,-7cm)}]
		\draw[thick, purple!30!black!70, postaction={decorate}] (3.5,8.5) .. controls (8,11) and (8,6) .. (11.5,8.5);
		\end{scope}
		
		\begin{scope}[decoration={markings,mark=at position 1 with {\arrow{>}}}, rotate around={-10:(3.5,1.5)}]
		\draw[thick, purple!30!black!70, postaction={decorate}] (3.5,1.5) .. controls (3.25,2) and (3.75,2) .. (3.5,2.5);
		\end{scope}
		
		\begin{scope}[decoration={markings,mark=at position 1 with {\arrow{>}}}, rotate around={55:(3.5,1.5)}]
		\draw[thick, purple!30!black!70, postaction={decorate}] (3.5,1.5) .. controls (3.25,2) and (3.75,2) .. (3.5,2.5);
		\end{scope}
		
		\begin{scope}[decoration={markings,mark=at position 1 with {\arrow{>}}}, rotate around={120:(3.5,1.5)}]
		\draw[thick, purple!30!black!70, postaction={decorate}] (3.5,1.5) .. controls (3.25,2) and (3.75,2) .. (3.5,2.5);
		\end{scope}
		
		\begin{scope}[decoration={markings,mark=at position 1 with {\arrow{>}}}, rotate around={240:(3.5,1.5)}]
		\draw[thick, purple!30!black!70, postaction={decorate}] (3.5,1.5) .. controls (3.25,2) and (3.75,2) .. (3.5,2.5);
		\end{scope}
	    
	    \filldraw[blue] (3.5,1.5) circle (3pt);
	    \filldraw[red] (11.5,1.5) circle (3pt);
	    
	    \node[label=:] (A2) at (-3,2) {\textcolor{purple!30!black!70}{Flows in the Lorentz}};
	    \node[label=:] (B2) at (-3,1) {\textcolor{purple!30!black!70}{invariant plane}};
	    \node[label=:] (C2) at (3.5,0.75) {\footnotesize \textcolor{blue}{UV}};
	    \node[label=:] (D2) at (11.5,2.25) {\footnotesize \textcolor{red}{IR}};
    \end{tikzpicture}
    \end{adjustbox}
    \caption{The figure emphasizes that the RG flow from 6d to 3d requires leaving the Lorentz-invariant RG plane by discontinuously turning on the curvature $\kappa\neq 0$.}
    \label{Twoflows}
    \end{figure}

\begin{itemize}

\item {\bf Explicit breaking of Lorentz invariance:} To trigger a flow across dimensions one need to break Lorentz invariance. The breaking of Lorentz invariance leads to a situation where standard quantum field theoretic methods are typically not well developed. In many cases, this breaking of Lorentz invariance is not continuous. That is, the parameter responsible cannot be continuously turned off. Holography, on the other hand,  is perfectly equipped to treat these cases. 

\item {\bf Beyond the relevant-deformation paradigm:} Flows across dimensions do not neatly fit the paradigm of perturbing a CFT by a relevant deformation. Consider compactifying the theory on a torus.  For a deformation $\delta g_{\mu\nu}$ of the background metric, it is natural to introduce the deformation as $S=S_{CFT}+ \lambda \int d^Dx \, \delta g_{\mu\nu} T^{\mu\nu}$.  However, for tori, $\delta g_{\mu\nu}$ vanishes as the metric remains locally flat; the deformation enters due to global identifications and cannot be captured by a local small deformation of the theory. A more drastic situation occurs in the space of deformations triggered by fluxes along the compactification manifold as some of these deformations are not even normalizable. 

\item {\bf Discontinuity from the Lorentz-preserving RG flows by fluxes:}  This property can the best described holographically. In particular, we show that one can never recover asymptotically $AdS_{D+1}$ in the UV. Fluxes cannot be continuously turned off. 
    
\item {\bf Discontinuity from the Lorentz-preserving RG flows by curvature:} Even in the absence of fluxes, the discontinuity can be sourced by  curvature in the compactification manifold. This is the example we study in this paper and the flow exists only for nonzero curvature denoted by $\kappa$ in Figure \ref{Twoflows}. We find a rich interplay between our results and studies of entanglement entropy with sub-regions containing curvature discussed in section~\ref{sec:anomalycoefs}.

\end{itemize}

There are clear obstacles to the existence of a $c$-function across dimensions; they can be clearly identified in the case of compatifications with fluxes. For instance, while there are methods to determine the central charge of the Lorentz-invariant fixed point, $c({\rm UV})$, there is currently no proper definition of $c(\widetilde{\rm UV})$.  Our goal in this manuscript is to continue explorations of potential entropic $c$-functions with emphasis on the role of curvature. 

%%%%%%%%%%%%%%%%%%%%%%%%%%%%%%%%%%%%%%%%%%
\subsection{Holographic RG flows across dimensions: fluxes and curvature}
The AdS/CFT correspondence can be used to analyze RG flows between a $D$-dimensional conformal field theory $\textrm{CFT}_D$ defined on $\mathbb{R}^{1,d-1} \times M_{D-d}$ in the UV regime and a $d$-dimensional conformal field theory $\textrm{CFT}_d$ defined on $\mathbb{R}^{1,d-1}$ in the IR regime, where the $\textrm{CFT}_d$ is obtained from the $\textrm{CFT}_D$ as the compactification on a ($D-d$)-dimensional compact space $M_{D-d}$. The holographic dual of this construction is given by a gravity solution interpolating between a ($D+1$)-dimensional asymptotically $\textrm{AdS}_{D+1}$ spacetime in the UV regime and an asymptotically $\textrm{AdS}_{d+1} \times M_{D-d}$ spacetime in the IR regime. The study of holographic RG flows across dimensions can be formulated in terms of the following metric
\begin{align}
	ds^2 = e^{2f(z)} \left( -dt^2 + dz^2 + dr^2 + r^2 d\Omega^2_{d-2} \right) + e^{2g(z)} \mathcal{\ell}^2 ds^2_{M_{D-d}},
\label{metricDd}
\end{align}
where the metric functions $f$ and $g$ of the holographic coordinate $z$ describe the interpolation between $\textrm{AdS}_{D+1}$ and $\textrm{AdS}_{d+1} \times M_{D-d}$. The Ansatz for this interpolation takes the following asymptotic form
\begin{subequations}
	\begin{align*}
	\textrm{UV regime} \ (z \rightarrow 0) :&& f(z) &\rightarrow \log(L_{\textrm{UV}}/z), &
	g(z) &\rightarrow \log(L_{\textrm{UV}}/z), \numberthis
    \label{asymptoticbehaviorDda}
    \\
	\textrm{IR regime} \ (z \rightarrow \infty) :&& f(z) &\rightarrow \log(L_{\textrm{IR}}/z), &
	g(z) &\rightarrow \tilde{g}_{\textrm{IR}}, \numberthis
	\label{asymptoticbehaviorDdb}
	\end{align*}
    \label{asymptoticbehaviorDd}%
\end{subequations}
where $L_{\textrm{UV}}$ is the radius of $\textrm{AdS}_{D+1}$, $L_{\textrm{IR}}$ is the radius of $\textrm{AdS}_{d+1}$, and $\tilde{g}_{\textrm{IR}}$ is a constant. Note that we have explicitly factored out the length dimension, $\ell$, of the compact manifold $M_{D-d}$.  There is a plethora of explicit solutions in this class (see  \cite{Maldacena:2000mw,Acharya:2000mu,Gauntlett:2000ng,Gauntlett:2001qs,Gauntlett:2001jj,Benini:2013cda,Benini:2015bwz,Bobev:2017uzs} for a partial list). 

Reference \cite{GonzalezLezcano:2022mcd}  proposed a systematic study of the behavior of holographic entanglement entropy along such holographic RG flows. In particular,  the holographic entanglement entropy associated with  entangling regions of the form $B^{d-1} \times M_{D-d}$, where $B^{d-1}$ is a $(d-1)$-dimensional ball of radius $R$ was studied. These are entangling regions that completely wrap the compactification manifold $M_{D-d}$, and can be given as a minimal surface in the bulk determined by the metric (\ref{metricDd}). The holographic entanglement entropy is determined by
\begin{align}
    S_{\textrm{EE}} \left( R; B^{d-1} \times M_{D-d}, \epsilon \right) &\nonumber\\
    &\kern-3cm=
    \frac{\textrm{vol}[S^{d-2}] \ \mathcal{\ell}^{D-d} \ \textrm{vol}[M_{D-d}]}{4 G^{(D+1)}_N}
    \min_{r(0)=R}
    \Bigg[ \displaystyle \int_\epsilon^{z_0} dz \ (r(z))^{d-2} \ e^{(d-1)\tilde{f}(z)}
    \sqrt{1+(r'(z))^2} \Bigg],
\label{holographicentanglemententropyDd}
\end{align}
where $G^{(D+1)}_N$ is the $(D+1)$-dimensional Newton's constant, $\epsilon$ is a UV cutoff in the bulk that imposes the restriction $z \geq \epsilon$, $z_0$ is the turning point of the minimal surface in the bulk, and the function $\tilde{f}$ stands for an effective warp factor defined by
\begin{align}
	\tilde{f}(z) = f(z) + \frac{D-d}{d-1}g(z).
\label{warpfactorDd}
\end{align}
Our interest in such surfaces is motivated by the results reported in \cite{Casini:2017roe,Casini:2017vbe} where monotonicity theorems were proven in dimensions two, three and four using the maximally spherically entangling region. We assume, as in those papers, that taking large values of $R$ provides insight into the IR fixed point of the RG flow.

As stated in Section \ref{introduction}, the holographic entanglement entropy (\ref{holographicentanglemententropyDd}) was analyzed in reference \cite{GonzalezLezcano:2022mcd} for holographic RG flows from

\begin{itemize}
    \item $\textrm{AdS}_5$ to $\textrm{AdS}_3 \times \Sigma_g$ ($D=4$, $d=2$) where $\Sigma_g$ is a genus $g$ Riemann surface with magnetic fluxes turned on.
    \item  $\textrm{AdS}_7$ to $\textrm{AdS}_3 \times \mathbb{H}_4$ ($D=6$, $d=2$) where $\mathbb{H}_4$ is a $4$-dimensional hyperbolic manifold. 
    \item $\textrm{AdS}_7$ to $\textrm{AdS}_5 \times T^2$ ($D=6$, $d=4$) where $T^2$ is the 2-torus with magnetic fluxes turned on.
\end{itemize}

In this paper we are interested in holographic RG flows from $\textrm{AdS}_7$ to $\textrm{AdS}_{4} \times \mathbb{H}_3$ ($D=6$, $d=3$) where $\mathbb{H}_3$ is a $3$-dimensional hyperbolic manifold. Our main interest is in considering a holographic RG flow across dimensions that is not directly supported by magnetic fluxes along the compactification manifold. One of our main findings is that the curvature of the compactification manifold alone, that is, without introducing fluxes, is sufficient to create a discontinuity in the space of RG flows. Figure \ref{Twoflows} describes the general situation. In subsequent sections we performed a detail study of entanglement entropy along this flow across dimensions.

Holographic RG flows have figured in other recent discussions. For example, reference \cite{Deddo:2022wxj} consider a surface that starts as the maximally spherical entangling surface in the UV and grows in the IR to wrap the compact space. Such analysis requires solving PDEs, given the lack of symmetries of the surface. Reference \cite{Deddo:2022wxj} also displayed an interesting interplay between various topology transitions; similar transitions were detected also in a related context in \cite{Fujita:2023bdk}. Another development was the consideration of flows across dimensions in the context of theories including higher-derivative corrections \cite{Deddo:2023pid}. An interesting application of RG flows across dimension was advanced recently in  \cite{Uhlemann:2021itz}
where the holographic duals for CFTs compactified on a Riemann surface $\Sigma_g$ with a twist were interpreted in the language of wedge holography with subsequent implications for Page curves. Emphasis of the implications of the Lorentz symmetry breaking flows for the counting of degrees of freedom has recently been discussed in \cite{Hoyos:2020zeg,Hoyos:2021vhl,Caceres:2023mqz}.
%%%%%%%%%%%%%%%%%%%%%%%%%%%%%%%%%%%%%%%%%%%%%%%%%%%%%%%%%%%%%%%%%%%%%%%%%%%%%%%%%%%%%%%%%%%%%%%%%%%%%%%%%%%%%%%%%%%%%%%%%%%%%%%%%%%%%%%%%%%%%%%%%%%%%%%%%%%%%%%%%%%%%%%%%%%%%%%%%%%%%%%
\section{Holographic RG flows from \texorpdfstring{$\textrm{AdS}_7$ to $\textrm{AdS}_4 \times \mathbb{H}_3$}{AdS7 to AdS4 x H3}}
\label{HolographicEE7to4}

The RG flow solution we are interested in has its origin in a consistent truncation of 11d supergravity to maximally supersymmetric 7d gauged supergravity on $S^4$. The Ansatz for the 11d metric is given by  
\cite{Pernici:1984xx,Nastase:1999cb,Nastase:1999kf}:

\begin{eqnarray}
    ds_{11d}^2&=& \Delta^{-\frac{2}{5}}ds_{7d}^2 +\frac{1}{m^2}\Delta^{\frac{4}{5}}\bigg[e^{4\lambda}DY^a DY^a+e^{-6\lambda} dY^i dY^i\bigg], \nonumber \\
    DY^a &=& dY^a + 2m\, B^{ab}\, Y^b, \nonumber \\
    \Delta^{-\frac{6}{5}}&=& e^{-4\lambda}Y^a Y^a + e^{6\lambda}Y^i Y^i,
\label{eq:11dans}
\end{eqnarray}
where $a=1,2,3, \, i =4,5$ and $(Y^a, Y^i)$ parametrize $S^4$: $Y^a Y^a + Y^i Y^i=1$. Notice that $m$ is the gauge coupling which determines the value of the radius of AdS$_7$. The gauge field $B^{ab}$ originates as a gauge field in $SO(5)$ but is restricted here to $SO(3)\in SO(5)$. We do not require the explicit form of the 11d four-form. From the above 11d Ansatz it becomes clear that the role of $\lambda$ is that of a squashing parameter for $S^4$ while in the 7d point of view it is simply a scalar field in the theory. 

In this manuscript we analyze RG flows from $\textrm{AdS}_7$ in the UV regime to $\textrm{AdS}_4 \times \mathbb{H}_3$ in the IR regime, where the compactification manifold $\mathbb{H}_3$ is a $3$-dimensional hyperbolic manifold. The BPS equations associated with compactifications on special Lagrangian (SLAG) $3$-cycles are given by \cite{Gauntlett:2000ng}
\begin{subequations}
\begin{align}
 e^{-f(z)}f'(z) + \frac{m}{10} \left[ 3e^{-4\lambda(z)} + 2e^{6\lambda(z)} \right]
               - \frac{3\kappa}{20m} e^{4\lambda(z) - 2g(z)} &= 0, \label{BPS-Eq-f} \\ 
	e^{-f(z)}g'(z) + \frac{m}{10} \left[ 3e^{-4\lambda(z)} + 2e^{6\lambda(z)} \right]
               + \frac{7\kappa}{20m} e^{4\lambda(z) - 2g(z)} &= 0, \label{BPS-Eq-g} \\ 
	e^{-f(z)}\lambda'(z) - \frac{m}{5} \left[ e^{6\lambda(z)} - e^{-4\lambda(z)} \right]
               - \frac{\kappa}{10m} e^{4\lambda(z) - 2g(z)} &= 0, \label{BPS-Eq-lambda}
\end{align}
\label{BPSequationsAdS7toAdS4z}%
\end{subequations}
where $f$ and $g$ are the metric functions in the parametrization (\ref{metricDd}), $\lambda$ is a scalar function, as given in (\ref{eq:11dans}), and $\kappa$ is the curvature of the SLAG $3$-cycle.

In the UV, the flow equations (\ref{BPSequationsAdS7toAdS4z}) admit an asymptotically AdS$_7$ solution
\begin{equation}
    f_{\mathrm{UV}}(z)\sim g_{\mathrm{UV}}(z)\sim\log\left(\fft{L_{\mathrm{UV}}}z\right),\qquad\lambda(z)\sim0,
\label{eq:UVfp}
\end{equation}
where $L_{\mathrm{UV}}=2/m$.  The flow away from the UV depends on the curvature, $\kappa$.  For $\kappa=0$, the UV is an AdS$_7$ fixed point, where (\ref{eq:UVfp}) is an exact solution.  In this case, turning on $\lambda$ drives a $\Delta=4$ relevant deformation.  However, the solution does not reach a stable IR fixed point and instead runs off into a singularity.  For $\kappa>0$, the curvature induces an RG flow away from the UV, and again the solution ends up singular in the IR.

Our main interest is in the negative curvature case, namely $\kappa<0$. Note that it is the curvature that induces the RG flow.  In this case, there is a stable IR fixed point corresponding to  $\textrm{AdS}_4 \times \mathbb{H}_3$ with the IR functions  \cite{Pernici:1984nw}
\begin{subequations}
	\begin{align*}
    f_{\textrm{IR}}(z) &= \log \left( \frac{e^{4\lambda_{\textrm{IR}}(z)}}{m} \frac{1}{z} \right) = \log \left( \frac{2^{2/5}}{m} \frac{1}{z} \right), \numberthis
    \label{IRfixedpointsolutionAdS7toAdS4za}
    \\ \\
	g_{\textrm{IR}}(z) &= \frac{1}{2} \log \left( \frac{-\kappa e^{8\lambda_{\textrm{IR}}(z)}}{2m^2} \right) = \frac{1}{2} \log \left( \frac{-\kappa}{2^{1/5} m^2} \right), \numberthis
    \label{IRfixedpointsolutionAdS7toAdS4zb}
    \\ \\
	\lambda_{\textrm{IR}}(z) &= \frac{1}{10} \log(2). \numberthis
	\label{IRfixedpointsolutionAdS7toAdS4zc}
	\end{align*}
    \label{IRfixedpointsolutionAdS7toAdS4z}%
\end{subequations}
This IR fixed point solution can be placed in the larger class of M5-branes wrapping supersymmetric cycles \cite{Gauntlett:2000ng}. We are interested in the supergravity solution that is dual to the superconformal field theory arising when an M5-brane wraps a negative curvature SLAG $3$-cycle $\mathbb{H}_3$. The above system of BPS flow equations, (\ref{BPSequationsAdS7toAdS4z}), admits a solution which smoothly flows from AdS$_7$ in the UV with asymptotic behavior (\ref{eq:UVfp}) to the $\textrm{AdS}_4 \times \mathbb{H}_3$ fixed point with (\ref{IRfixedpointsolutionAdS7toAdS4z}) in the IR.  In what follows, we will discuss details of this holographic flow; some existence explorations were  presented in \cite{Gauntlett:2000ng}.

The solutions of the BPS equations are crucial for the construction of the holographic entanglement entropy. In this section  we use analytic expansions to describe the solution to the BPS equations (\ref{BPSequationsAdS7toAdS4z})  in the UV and IR regions. We then proceed to construct a numerical solution interpolating between the two regions.

%%%%%%%%%%%%%%%%%%%%%%%%%%%%%%%%%%%%%%%%%%%%%%%%%%%%%%%%%%%%%%%%%%%%%%%%%%%%%%%%%%%%%%%%%%%%%%%%%%%%%%%%%%%%%%%%%%%%%%%%%%%%%%%%%%%%%%%%%%%%%%%%%%%%%%%%%%%%%
\subsection{The UV regime}
\label{BPSUV}

We start by exploring the UV behavior of the BPS equations.  Ignoring back-reaction, we can start with the UV geometry, $e^f=e^g=L_{\mathrm{UV}}/z$.  In this case, the linearized $\lambda$ equation admits a solution of the form
\begin{equation}
    \lambda(z)=  - \frac{\kappa}{10m^2} \left( \frac{z}{L_\textrm{UV}} \right)^2 
    +A\left( \frac{z}{L_\textrm{UV}} \right)^4,
\label{eq:lamasy}
\end{equation}
where $A$ parametrizes the strength of the relevant scalar deformation.  This demonstrates that the flow is driven at $\mathcal O(z^2)$ by the (negative) curvature, while the scalar deformation first enters at $\mathcal O(z^4)$.  Working to non-linear order, and with back-reaction, we see that the $f(z)$, $g(z)$ and $\lambda(z)$ functions can be expanded beyond the UV logs, (\ref{eq:UVfp}), as a series in even powers of $z$, with the inclusion of log terms starting at $\mathcal O(z^4)$.  We thus expand in the UV regime ($z \rightarrow 0$) using the following Ansatz:
\begin{align*}
	f_{\textrm{UV}}(z) &= -\log \left( z/L_{\textrm{UV}} \right) 
    + \sum_{i=1}^{\infty} \sum_{j=0}^{i-1} F_{i,j} \left( z/L_{\textrm{UV}} \right)^{2i} 
    \left[ \log \left( z/L_{\textrm{UV}} \right) \right]^j, \\ \\
	g_{\textrm{UV}}(z) &= -\log \left( z/L_{\textrm{UV}} \right) 
    + \sum_{i=1}^{\infty} \sum_{j=0}^{i-1} G_{i,j} \left( z/L_{\textrm{UV}} \right)^{2i} 
    \left[ \log \left( z/L_{\textrm{UV}} \right) \right]^j, \\ \\
	\lambda_{\textrm{UV}}(z) &= 
    \sum_{i=1}^{\infty} \sum_{j=0}^{i-1} \Lambda_{i,j} \left( z/L_{\textrm{UV}} \right)^{2i} 
    \left[ \log \left( z/L_{\textrm{UV}} \right) \right]^j. \numberthis
	\label{expansionoffunctionsAdS7toAdS4UV}
\end{align*}
The coefficients $F_{i,j}$, $G_{i,j}$ and $\Lambda_{i,j}$ can be found by perturbatively solving the BPS equations (\ref{BPSequationsAdS7toAdS4z}) with the Ansatz (\ref{expansionoffunctionsAdS7toAdS4UV}).  Since we have fixed the UV asymptotics according to (\ref{eq:UVfp}), we find that the solution has only one free parameter, namely the coefficient $\Lambda_{2,0}$, corresponding to the relevant deformation parameter, $A$, in (\ref{eq:lamasy}).

The perturbative solution was obtained in {\it Mathematica} up to order four in the variable $z$.  The resulting metric functions take the form
\begin{align*}
    f_{\textrm{UV}}(z) &= -\log \left( z/L_{\textrm{UV}} \right) 
    + \frac{\kappa}{10m^2} \left( \frac{z}{L_\textrm{UV}} \right)^2 
    + \frac{\kappa^2}{200m^4} \left( \frac{z}{L_\textrm{UV}} \right)^4+\cdots, \\ \\
    g_{\textrm{UV}}(z) &= -\log \left( z/L_{\textrm{UV}} \right) 
    - \frac{2\kappa}{5m^2} \left( \frac{z}{L_\textrm{UV}} \right)^2
    - \frac{3 \kappa^2}{25 m^4} \left( \frac{z}{L_\textrm{UV}} \right)^4+\cdots. \numberthis
    \label{metricfunctionsorderfourAdS7toAdS4UV}
\end{align*}
Note that the free parameter $\Lambda_{2,0}$ only enters the metric at higher order.  It, of course, enters directly in the scalar function, which takes the form
\begin{equation}
    \lambda_{\textrm{UV}}(z) = 
    - \frac{\kappa}{10m^2} \left( \frac{z}{L_\textrm{UV}} \right)^2 
    + \left[ \Lambda_{2,0} + \frac{\kappa^2}{10m^4} \log\left( \frac{z}{L_\textrm{UV}} \right) \right] \left( \frac{z}{L_\textrm{UV}} \right)^4+\cdots.
    \label{scalarfieldorderfourAdS7toAdS4UV}
\end{equation}
The initial conditions are such that, as $z \rightarrow 0$, the metric functions $f_{\textrm{UV}}$ and $g_{\textrm{UV}}$ match the asymptotic UV behavior (\ref{asymptoticbehaviorDd}) while the scalar function $\lambda_{\textrm{UV}}$ vanishes.

We note that this UV expansion is valid for any value of the curvature $\kappa$.  However, the form of the flow away from the UV depends on both $\kappa$ and $\Lambda_{2,0}$.  For $\kappa>0$, the solution always flows to a singular IR, with either $\lambda$ running away to $+\infty$ or $-\infty$, depending on whether $\Lambda_{2,0}$ is greater than or less than some critical value dependent on $\kappa$.  For $\kappa=0$, there is a fixed point AdS$_7$ solution with $\Lambda_{2,0}=0$, while positive $\Lambda_{2,0}$ induces a singular flow to $\lambda\to+\infty$ and negative $\Lambda_{2,0}$ generates a flow to $\lambda\to-\infty$.  The configuration we are interested in is that of $\kappa<0$, where there exists a stable IR fixed point, (\ref{IRfixedpointsolutionAdS7toAdS4z}).  For a fixed $\kappa<0$, there is a critical value of $\Lambda_{2,0}\equiv\Lambda_*$, which can be determined numerically, such that the flow from the UV lands on the IR fixed point.  For $\Lambda_{2,0}>\Lambda_*$, the solution will run off to $\lambda\to+\infty$, and for $\Lambda_{2,0}<\Lambda_*$, the solution will run off to $\lambda\to-\infty$.

%%%%%%%%%%%%%%%%%%%%%%%%%%%%%%%%%%%%%%%%%%%%%%%%%%%%%%%%%%%%%%%%%%%%%%%%%%%%%%%%%%%%%%%%%%%%%%%%%%%%%%%%%%%%%%%%%%%%%%%%%%%%%%%%%%%%%%%%%%%%%%%%%%%%%%%%%%%%%
\subsection{The IR regime}
\label{BPSIR}

Assuming the flow ends up at the IR fixed point (which requires $\kappa<0$), we can likewise perform a perturbative expansion of the functions $f$, $g$ and $\lambda$ in the IR regime ($z \rightarrow \infty$) around the fixed point solution (\ref{IRfixedpointsolutionAdS7toAdS4z}).  Linearizing around the IR fixed point, we take
\begin{align*}
    f_{\textrm{IR}}(z) &= \log \left( L_{\textrm{IR}}/z \right) + \epsilon\, f_1(z), \\ \\
    g_{\textrm{IR}}(z) &= g_0 + \epsilon\, g_1(z), \\ \\
    \lambda_{\textrm{IR}}(z) &= \lambda_0 + \epsilon\, \lambda_1(z), \numberthis
    \label{expansionoffunctionsAdS7toAdS4IR}
\end{align*}
and work up to linear order in $\epsilon$. At order zero, we find
\begin{align}
    L_{\textrm{IR}} = \frac{2^{2/5}}{m}, \qquad g_0 = \frac{1}{2} \log \left( \frac{-\kappa}{2^{1/5} m^2} \right), \qquad \lambda_0 &= \frac{1}{10} \log(2),
    \label{coefficientsAdS7toAdS4IR0}
\end{align}
which agrees with the IR fixed point solution (\ref{IRfixedpointsolutionAdS7toAdS4z}). At order one, the first-order flow equations admit three eigenvectors, and the perturbative solution can be written as
\begin{align*}
    %L_{\textrm{IR}} &= \frac{2^{2/5}}{m}, \\ \\
    %g_0 &= \frac{1}{2} \log \left( \frac{1}{2^{1/5} m^2} \right), \\ \\
    %\lambda_0 &= \frac{1}{10} \log(2), \\ \\
    %%%%%%%%%%%%%%%%%%%%%%%%%%%%%%%%%%%%%%%%%%%%%%%%%%%%%%%%%%%%%%%%%%%%%%%%%%%%%%%%%%%%%
	f_1(z) &= c_1 \left( \frac{z}{L_{\textrm{IR}}} \right)^{\frac{1}{2} - \frac{\sqrt{17}}{2}} + c_2 \left( \frac{z}{L_{\textrm{IR}}} \right)^{\frac{1}{2} + \frac{\sqrt{17}}{2}} + \frac{c_3}{z}, \\ \\
    g_1(z) &= - \frac{2}{3} c_1 \left( \frac{z}{L_{\textrm{IR}}} \right)^{\frac{1}{2} - \frac{\sqrt{17}}{2}} - \frac{2}{3} c_2 \left( \frac{z}{L_{\textrm{IR}}} \right)^{\frac{1}{2} + \frac{\sqrt{17}}{2}}, \\ \\
    %%%%%%%%%%%%%%%%%%%%%%%%%%%%%%%%%%%%%%%%%%%%%%%%%%%%%%%%%%%%%%%%%%%%%%%%%%%%%%%%%%%%%
    \lambda_1 (z) &= c_1 \left( - \frac{19}{24} + \frac{5}{24} \sqrt{17} \right) \left( \frac{z}{L_{\textrm{IR}}} \right)^{\frac{1}{2} - \frac{\sqrt{17}}{2}} +
    c_2 \left( - \frac{19}{24} - \frac{5}{24} \sqrt{17} \right) \left( \frac{z}{L_{\textrm{IR}}} \right)^{\frac{1}{2} + \frac{\sqrt{17}}{2}},
    \numberthis
	\label{coefficientsAdS7toAdS4IR1}
\end{align*}
where there are three free parameters, namely $c_1$, $c_2$ and $c_3$. The solution associated with $c_2$ is a relevant deformation in the IR as it diverges when $z \rightarrow \infty$, while the solution associated with $c_3$ is marginal, corresponding to a shift $z \rightarrow z + c_3$.  Since we are interested in flows terminating at this IR fixed point, we focus on the irrelevant $c_1$ deformation.  Thus we take
\begin{align*}
    f_{\textrm{IR}}(z) &= \log \left( \frac{L_{\textrm{IR}}}{z} \right) +  c_1 \left( \frac{z}{L_{\textrm{IR}}} \right)^{\frac{1}{2} - \frac{\sqrt{17}}{2}}+\cdots, \\ \\
    g_{\textrm{IR}}(z) &= \frac{1}{2} \log \left( \frac{-\kappa}{2^{1/5} m^2} \right) -  \frac{2}{3} c_1 \left( \frac{z}{L_{\textrm{IR}}} \right)^{\frac{1}{2} - \frac{\sqrt{17}}{2}} +\cdots, \numberthis
    \label{metricfunctionsorderoneAdS7toAdS4IR}
\end{align*}
along with the scalar function
\begin{equation}
    \lambda_{\textrm{IR}}(z) = \frac{1}{10} \log(2) + c_1 \left( - \frac{19}{24} + \frac{5}{24} \sqrt{17} \right) \left( \frac{z}{L_{\textrm{IR}}} \right)^{\frac{1}{2} - \frac{\sqrt{17}}{2}}+\cdots.
\label{scalarfunctionorderoneAdS7toAdS4IR}
\end{equation}
Note that for the condition $z \rightarrow \infty$ the metric functions $f_{\textrm{IR}}$ and $g_{\textrm{IR}}$ match the asymptotic IR behavior (\ref{asymptoticbehaviorDd}) while the scalar function $\lambda_{\textrm{IR}}$ behaves like the constant $\frac{1}{10} \log(2)$.

%%%%%%%%%%%%%%%%%%%%%%%%%%%%%%%%%%%%%%%%%%%%%%%%%%%%%%%%%%%%%%%%%%%%%%%%%%%%%%%%%%%%%%%%%%%%%%%%%%%%%%%%%%%%%%%%%%%%%%%%%%%%%%%%%%%%%%%%%%%%%%%%%%%%%%%%%%%%%
\subsection{Complete numerical solution}
\label{BPSnumerical}

Given the asymptotic UV and IR solutions, we wish to construct numerical flows connecting them together.  In the UV, the solution (\ref{metricfunctionsorderfourAdS7toAdS4UV}) and (\ref{scalarfieldorderfourAdS7toAdS4UV}) depends on the deformation parameter $\Lambda_{2,0}$, which needs to be adjusted, \textit{e.g.} by a shooting method, in order to land on the IR fixed point.  On the other hand, while the IR solution
(\ref{metricfunctionsorderoneAdS7toAdS4IR}) and (\ref{scalarfunctionorderoneAdS7toAdS4IR}) also depends on one parameter, $c_1$, integrating towards the UV will always land on the asymptotic AdS$_7$ solution (provided $c_1$ is chosen to be negative).  Without adjusting $c_3$, the resulting UV solution may not terminate at $z=0$, but instead at some constant value $z_0$.  Nevertheless, this shift in $z$ can be easily accommodated, and hence no shooting is required when integrating up from the IR to the UV.  This was the approach taken in \cite{GonzalezLezcano:2022mcd}.

\begin{figure}[ht]
\centering
    \begin{subfigure}[b]{0.45\textwidth}
    \centering
    \includegraphics[scale=0.78]{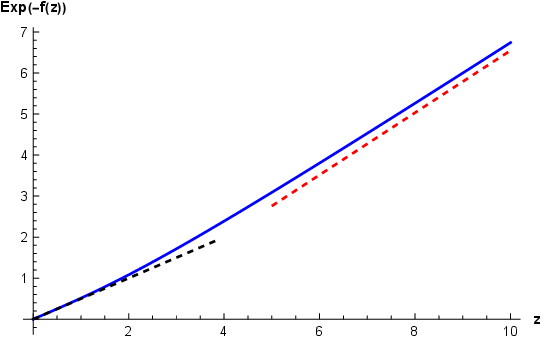}
    \caption{Numerical solution of the metric function $f$ (blue line) in the interval $z \in [1 \times 10^{-5}, 10]$. It approaches the behavior (\ref{asymptoticbehaviorDda}) in the UV regime (dashed black line) and the IR fixed point solution (\ref{IRfixedpointsolutionAdS7toAdS4za}) in the IR regime (dashed red line).}
    \label{BPSnumericalsolutionAdS7toAdS4aversion2}
    \end{subfigure}
    \hfill
    \begin{subfigure}[b]{0.45\textwidth}
    \centering
    \includegraphics[scale=0.78]{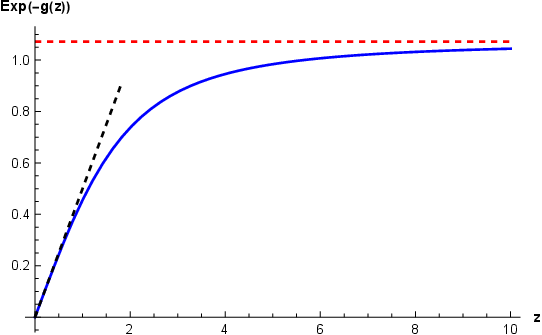}
    \caption{Numerical solution of the metric function $g$ (blue line) in the interval $z \in [1 \times 10^{-5}, 10]$. It approaches the behavior (\ref{asymptoticbehaviorDda}) in the UV regime (dashed black line) and the IR fixed point solution (\ref{IRfixedpointsolutionAdS7toAdS4zb}) in the IR regime (dashed red line).}
    \label{BPSnumericalsolutionAdS7toAdS4bversion2}
    \end{subfigure} \\ \vspace{0.5cm}
    \begin{subfigure}[b]{0.45\textwidth}
    \centering
    \includegraphics[scale=0.78]{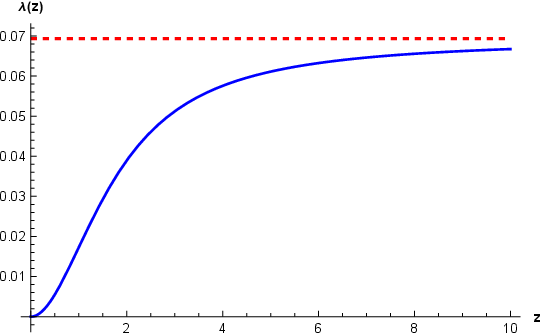}
    \caption{Numerical solution of the scalar function $\lambda$ (blue line) in the interval $z \in [1 \times 10^{-5}, 10]$. It approaches the IR fixed point solution (\ref{IRfixedpointsolutionAdS7toAdS4zc}) in the IR regime (dashed red line).}
    \label{BPSnumericalsolutionAdS7toAdS4cversion2}
    \end{subfigure}
\caption{Numerical solution of the BPS equations associated with holographic RG flows from $\textrm{AdS}_7$ to $\textrm{AdS}_4 \times \mathbb{H}_3$.  The numerical solution is constructed with $m=1$ and $\kappa=-1$ (\textit{i.e.}~$L_{\mathrm{UV}}=2$).}
\label{BPSnumericalsolutionAdS7toAdS4version2}
\end{figure}

While starting from the IR has the benefit of avoiding having to shoot, here we nevertheless will start in the UV and perform a shooting method to adjust $\Lambda_{2,0}$ to flow to the IR.  The reason for this is that it enables us to simultaneously construct minimal entangling surfaces anchored in the UV.  For the actual numerical integration, we used NDSolve in {\it Mathematica} by setting $\textrm{Method} \rightarrow \textrm{``StiffnessSwitching''}$, $\textrm{WorkingPrecision} \rightarrow 50$ and $\textrm{MaxSteps} \rightarrow 10,000$.  For all of the numerical work, we take $m=1$, corresponding to $L_{\mathrm{UV}}=2$.  For $\kappa/m^2=-1$ (\textit{i.e.}~$\kappa=-1$ in $m=1$ units), the parameter $\Lambda_{2,0}$ was found to be $\Lambda_{2,0}=\Lambda_* \approx -0.056568787$ by using a shooting method and by taking higher order terms of expressions (\ref{metricfunctionsorderfourAdS7toAdS4UV}) and (\ref{scalarfieldorderfourAdS7toAdS4UV}) into account. The numerical solution is shown in figure \ref{BPSnumericalsolutionAdS7toAdS4version2}.
Note that the plots of the functions $\exp{(-f)}$ and $\exp{(-g)}$ in figures \ref{BPSnumericalsolutionAdS7toAdS4aversion2} and \ref{BPSnumericalsolutionAdS7toAdS4bversion2}, respectively, match both the asymptotic behaviors (\ref{asymptoticbehaviorDd}) and the IR fixed point solution (\ref{IRfixedpointsolutionAdS7toAdS4z}). In figure \ref{BPSnumericalsolutionAdS7toAdS4aversion2}, it was emphasized that the function $\exp{(-f)}$ has two leading slopes, one for the UV regime (\ref{asymptoticbehaviorDda}) and another one for the IR regime (\ref{IRfixedpointsolutionAdS7toAdS4za}) (see also (\ref{asymptoticbehaviorDdb})). In figure \ref{BPSnumericalsolutionAdS7toAdS4bversion2}, it can be seen that the UV behavior of the function $\exp{(-g)}$ agrees with (\ref{asymptoticbehaviorDda}) while the IR behavior agrees with (\ref{IRfixedpointsolutionAdS7toAdS4zb}) (see also (\ref{asymptoticbehaviorDdb})). Furthermore, it can be seen in figure \ref{BPSnumericalsolutionAdS7toAdS4cversion2} that the scalar function $\lambda$ vanishes in the UV regime while it tends to the constant value of expression (\ref{IRfixedpointsolutionAdS7toAdS4zc}) in the IR regime.

%%%%%%%%%%%%%%%%%%%%%%%%%%%%%%%%%%%%%%%%%%%%%%%%%%%%%%%%%%%%%%%%%%%%%%%%%%%%%%%%%%%%%%%%%%%%%%%%%%%%%%%%%%%%%%%%%%%%%%%%%%%%%%%%%%%%%%%%%%%%%%%%%%%%%%%%%%%%%
\section{Holographic entanglement entropy along the RG flow}
\label{HEE}

After numerically constructing the holographic RG flow from $\textrm{AdS}_7$ to $\textrm{AdS}_4 \times \mathbb{H}_3$, we now turn to the computation of the holographic entanglement entropy.  Inserting $D=6$ and $d=3$ into the expression for the minimal surface, (\ref{holographicentanglemententropyDd}), we have%
\footnote{The volume of the ($n-1$)-dimensional sphere $S^{n-1}$ is given by $\textrm{vol}[S^{n-1}] = 2 \pi^{n/2}/\Gamma \left( n/2 \right)$.}
\begin{align*}
    S_{\textrm{EE}} \left( R; B^2 \times \mathbb{H}_3, \epsilon \right)
    =
    \frac{2 \pi \mathcal{\ell}^3 \textrm{vol}[\mathbb{H}_3]}{4 G^{(7)}_N}
    \min_{r(0)=R}
    \Bigg[ \displaystyle \int_\epsilon^{z_0} dz \ r(z) \ e^{2\tilde{f}(z)}
    \sqrt{1+(r'(z))^2} \Bigg], \numberthis
    \label{holographicentanglemententropyAdS7toAdS4}
\end{align*}
where the IR warp factor (\ref{warpfactorDd}) has the form
\begin{align}
	\tilde{f}(z) = f(z) + \frac{3}{2}g(z),
\label{warpfactorAdS7toAdS4}
\end{align}
and $r(z)$ is the embedding function that parameterizes the minimal surface in the bulk.

Varying the action, (\ref{holographicentanglemententropyAdS7toAdS4}), with respect to $r(z)$ gives the second order ordinary differential equation%
\footnote{The RG flows from $\textrm{AdS}_7$ to $\textrm{AdS}_3 \times \mathbb{H}_4$ and from $\textrm{AdS}_5$ to $\textrm{AdS}_3 \times \Sigma_g$ studied in reference \cite{GonzalezLezcano:2022mcd} led to a more tractable differential equation as well as a first integral, as there was no explicit dependence on the function $r$ in the Lagrangian.}
\begin{align}
	r(z) \left[ 2 \tilde{f}'(z) r'(z) \left( 1 + (r'(z))^2 \right) + r''(z) \right] - (r'(z))^2 - 1 = 0,
\label{differentialequationAdS7toAdS4}
\end{align}
for the profile of the surface.  Since $\tilde f(z)$ is only known numerically along the flow, the solution for the minimal surface will have to be constructed numerically.  This will be performed in section~\ref{wholeentanglemententropy}. But first, as a warm-up, we compute the UV and IR structures of the holographic entanglement entropy (\ref{holographicentanglemententropyAdS7toAdS4}) in sections \ref{UVentanglemententropy} and \ref{IRentanglemententropy}, respectively.

%%%%%%%%%%%%%%%%%%%%%%%%%%%%%%%%%%%%%%%%%%%%%%%%%%%%%%%%%%%%%%%%%%%%%%%%%%%%%%%%%%%%%%%%%%%%%%%%%%%%%%%%%%%%%%%%%%%%%%%%%%%%%%%%%%%%%%%%%%%%%%%%%%%%%%%%%%%%%
\subsection{UV regime (small entangling region)}
\label{UVentanglemententropy}

The small entangling region is specified by the conditions $z_0 \ll L_{\textrm{UV}}$ and $R \ll L_{\textrm{UV}}$, where $L_{\textrm{UV}}$ is the radius of $\textrm{AdS}_7$.  In this case, to a good approximation, we can use the perturbative expansion of the metric functions $f_{UV}$ and $g_{UV}$ given in  (\ref{metricfunctionsorderfourAdS7toAdS4UV})%
\footnote{In principle, we can expand the metric functions to arbitrary perturbative order.  However, the expansion to $\mathcal O(z^4)$ in (\ref{metricfunctionsorderfourAdS7toAdS4UV}) is sufficient to identify the UV divergences of the entanglement entropy.}.
In this way, the differential equation determining the surface, (\ref{differentialequationAdS7toAdS4}), can be written analytically, although the full solution for $r(z)$ will still need to be determined numerically.  In particular, we will find analytical expressions for the surface in the regions $z \rightarrow 0$ and $z \rightarrow z_0$, and then we will join them together by using a numerical solution.

%%%%%%%%%%%%%%%%%%%%%%%%%%%%%%%%%%%%%%%%%%%%%%%%%%%%%%%%%%%%%%%%%%%%%%%%%%%%%%%%%%%%%%%%%%%%%%%%%%%%%%%%%%%%%%%%%%%%%%%%%%%%%%%%%%%%%%%%%%%%%%%%%%%%%%%%%%%%%
\subsubsection{Analytical expression around \texorpdfstring{$z \rightarrow 0$}{}.}
\label{UVentanglemententropy-a}

In the region were $z \rightarrow 0$ the holographic entanglement entropy (\ref{holographicentanglemententropyAdS7toAdS4}) can be written as
    \begin{align*}
    S^{\textrm{EE}}_{\textrm{UV}_{z \rightarrow 0}} \left( R; B^2 \times \mathbb{H}_3, \epsilon \right)
    =
    \frac{2 \pi \mathcal{\ell}^3 \textrm{vol}[\mathbb{H}_3]}{4 G^{(7)}_N}
    \min_{r(0)=R}
    \Bigg[ \displaystyle \int_\epsilon^{z_0} dz \ A_{\textrm{UV}_{z \rightarrow 0}} (z,r(z),r'(z)) \Bigg], \numberthis
    \label{holographicentanglemententropyAdS7toAdS4preUV}
    \end{align*}
where the Lagrangian $A_{\textrm{UV}_{z \rightarrow 0}}$ takes the form
\begin{align}
    A_{\textrm{UV}_{z \rightarrow 0}} (z,r(z),r'(z)) 
    = r(z) \left[ e^{2f_{\textrm{UV}}(z) + 3g_{\textrm{UV}}(z)} \right] \sqrt{1+(r'(z))^2}.
    \label{areaAdS7toAdS4UV}
\end{align}
So long as we restrict the minimal surface to be fully contained in the UV, we may take the metric functions $f_{\textrm{UV}}$ and $g_{\textrm{UV}}$ to be given by the asymptotic solution, (\ref{metricfunctionsorderfourAdS7toAdS4UV}). The associated variational principle with respect to the function $r(z)$  leads to the second order ordinary differential equation
\begin{align*}
     r(z)
    \Big[ 400 + 40 z^2 \kappa + 7 z^4 \kappa^2 \Big] 
    \Big[ r'(z) + \left( r'(z) \right)^3 & \Big]
    \\
    + 80  z 
    \Big[ 1 + (r'(z))^2 - r(z) r''(z) & \Big] = 0. \numberthis
    \label{differentialequationAdS7toAdS4UV}
\end{align*}
This equation is to be solved subject to the boundary condition $\displaystyle \lim_{z \rightarrow 0} r(z)=R$.  Although we have not found an exact solution to (\ref{differentialequationAdS7toAdS4UV}), it can be solved perturbatively around $z \rightarrow 0$ by using the following Ansatz
\begin{align}
    r_{\textrm{UV}}(z) = R + \sum_{i=1}^{\infty} \sum_{j=0}^{i-1} a_{i,j} \ z^{2i} \ [\log (z)]^j.
    \label{AnsatzrUV}
\end{align}
The solution to sixth order in $z$ is given by 
\begin{align*}
    r_{\textrm{UV}}(z) = R
    & + \left( \frac{- 1}{8 R} \right) z^2 + \left( \frac{-7 + 8 R^2 \kappa}{512 R^3} \right) z^4 \\
    & + \left[ a_{3,0} + \left( \frac{45  - 80  R^2 \kappa + 32 R^4 \kappa^2}{20480  R^5} \right) \log(z) \right] z^6+\cdots. \numberthis
    \label{parameterizingfunctionrUV}
\end{align*}
Note that this solution depends on one free parameter, $a_{3,0}$, which needs to be adjusted so that the minimal surface caps off smoothly at $z_0$.  The terms beyond $\mathcal O(z^6)$ will depend on higher order terms in the metric functions, (\ref{metricfunctionsorderfourAdS7toAdS4UV}).  However, they will not contribute to the leading order divergences.

The asymptotic expansion, (\ref{parameterizingfunctionrUV}), is sufficient to determine the leading divergences of the holographic entanglement entropy
\begin{align*}
    S^{\textrm{EE}}_{\textrm{UV}_{z \rightarrow 0}} \left( R; B^2 \times \mathbb{H}_3, \epsilon \right)
    & =
    \frac{2 \pi \mathcal{\ell}^3 \textrm{vol}[\mathbb{H}_3]}{4 G^{(7)}_N}
    \int_\epsilon^{z_0} dz 
    \ r_{\textrm{UV}}(z) \left[ e^{2f_{\textrm{UV}}(z) + 3g_{\textrm{UV}}(z)} \right] \sqrt{1+(r_{\textrm{UV}}'(z))^2} \\
    %%%%%%%%%%%%%%%%%%%%%%%%%%%%%%
    & =
    \frac{2 \pi \mathcal{\ell}^3 \textrm{vol}[\mathbb{H}_3]L_{\mathrm{UV}^5}}{4 G^{(7)}_N}
    \int_\epsilon^{z_0}dz\biggl(\fft{R}{z^5}-\fft{3+8\kappa R^2}{32Rz^3}\\
    &\kern13em-\fft{3(15-80\kappa R^2-32\kappa^2R^4)}{10240R^3z}+\mathcal O(z)\biggr) \\    
    %%%%%%%%%%%%%%%%%%%%%%%%%%%%%%
    & =
    \frac{2 \pi \mathcal{\ell}^3 \textrm{vol}[\mathbb{H}_3]L_{\mathrm{UV}}^5}{4 G^{(7)}_N R^3}
    \Bigg[\fft14\left(\fft{R}\epsilon\right)^4 + A\left(\fft{R}\epsilon\right)^2    + B
    \log \left( \frac{R}{\epsilon} \right)
    + \mathcal{O}(\epsilon^0) \Bigg], \numberthis
\label{holographicentanglemententropyAdS7toAdS4UVrofz}
\end{align*}
where the coefficients $A$ and $B$ are given by
\begin{align*}
    A &= -\left( \frac{3}{64}+\frac{\kappa R^2}{8}\right), \\
    B &= -\frac{3}{10240}
    \left(15-80\kappa R^2-32 \kappa^2R^4\right).
    \label{holographicentanglemententropyAdS7toAdS4UVcoefficientsAB}
    \numberthis
\end{align*}
The $1/\epsilon^4$, $1/\epsilon^2$ and logarithmic divergences in (\ref{holographicentanglemententropyAdS7toAdS4UVrofz}) are typical of $\textrm{AdS}_7$. However, note that the $A$ and $B$ coefficients include subleading contributions to the holographic entanglement entropy that depend on the curvature $\kappa$ of the compact manifold, measured in units of the radius $R$ of the entangling region.

We are, of course, interested in the finite $\mathcal O(\epsilon^0)$ contribution to the entanglement entropy.  However, this cannot be obtained directly from the perturbative expansion of the minimal surface around $z=0$, as the expansion of $r(z)$ breaks down as $z\to z_0$.  In particular, assuming a smooth capoff point at $z_0$, the profile of the minimal surface behaves according to $r^2\sim(z_0-z)$ near $z_0$.  Hence $r(z)$ has a branch cut singularity, and we will have to resort to numerical integration to obtain the finite term as a function of the radius $R$.

\subsubsection{Analytical expression around \texorpdfstring{$z \rightarrow z_0$}{}}
\label{UVentanglemententropy-b}

To facilitate the numerical integration for the profile of the entangling surface, we consider expanding around the capoff point, $z=z_0$.  In this case, since $r'(z_0)\to\infty$, we turn the problem around to consider $z$ as a function of $r$ satisfying the conditions $z(0)=z_0$ and $z'(0)=0$.  The holographic entanglement entropy (\ref{holographicentanglemententropyAdS7toAdS4}) is then written as
\begin{equation}
    S_{\textrm{EE}} \left( R; B^2 \times \mathbb{H}_3 \right)
    =
    \frac{2 \pi \mathcal{\ell}^3 \textrm{vol}[\mathbb{H}_3]}{4 G^{(7)}_N}
    \min_{z(0)=z_0}
    \Bigg[ \displaystyle \int_0^{R} dr \ r \ e^{2\tilde{f}(z(r))}
    \sqrt{1 + (z'(r))^2} \Bigg],
\label{holographicentanglemententropyAdS7toAdS4preUVzofr}
\end{equation}
where the upper limit, $R$, will have to be regulated to handle the UV divergence.   Noting that we are still working in the UV regime of the metric, we can write this as
\begin{align*}
    S^{\textrm{EE}}_{\textrm{UV}_{r \rightarrow 0}} \left( R; B^2 \times \mathbb{H}_3 \right)
    =
    \frac{2 \pi \mathcal{\ell}^3 \textrm{vol}[\mathbb{H}_3]}{4 G^{(7)}_N}
    \min_{z(0)=z_0}
    \Bigg[ \displaystyle \int_0^{R} dr \ A_{\textrm{UV}_{r \rightarrow 0}} (r, z(r), z'(r)) \Bigg],
    \numberthis
    \label{holographicentanglemententropyAdS7toAdS4preUVzofrddd}
\end{align*}
where the Lagrangian $A_{\textrm{UV}_{r \rightarrow 0}}$ is
\begin{align}
    A_{\textrm{UV}_{r \rightarrow 0}} (r, z(r), z'(r)) =  r \left[ e^{2f_{\textrm{UV}}(z(r)) + 3g_{\textrm{UV}}(z(r))} \right]
    \sqrt{1 + (z'(r))^2},
\end{align}
where the metric functions $f_{\textrm{UV}}$ and $g_{\textrm{UV}}$ are given by (\ref{metricfunctionsorderfourAdS7toAdS4UV}). The associated variational principle with respect to the function $z(r)$ leads to the following non-linear second order differential equation
\begin{align*}
    r\left( 1 + (z'(r))^2 \right) 
    \Big[ 400 + 40 \kappa (z(r))^2 + 7 \kappa^2 (z(r))^4 \Big]& \\
    + 80 z(r) \Big[ z'(r) + (z'(r))^3 + r z''(r) \Big]& = 0. \numberthis
\label{differentialequationAdS7toAdS4UVzofr}
\end{align*}

Recall that the embedding function $z(r)$ must satisfy the conditions $z(0)=z_0$ and $z'(0)=0$. This further implies that $z(r)$ is an even function of $r$, so equation (\ref{differentialequationAdS7toAdS4UVzofr}) can be solved around $r \rightarrow 0$ using the following Ansatz
\begin{align}
    %z_{\textrm{UV}}(r) = z_0 + z_1 r + z_2 r^2 + z_3 r^3 + z_4 r^4,
    z_{\textrm{UV}}(r) = z_0 + \sum_{i=1}^{\infty} z_{2i} r^{2i}.
\label{AnsatzzUV}
\end{align}
Solving perturbatively up to $\mathcal O(r^4)$ gives
\begin{equation}
    z_{\textrm{UV}}(r) =z_0+z_2r^2+z_4r^4+\mathcal O(r^6),
\label{parameterizingfunctionzUV}
\end{equation}
where
\begin{align*}
    z_2 &= -\frac{5}{4z_0}\left(1+\frac{1}{10}\kappa z_0^2+\frac{7}{400}\kappa^2z_0^4 +\mathcal O(\kappa^3)\right), \\
    z_4 &= -\fft{175}{128z_0^3}\left(1+\fft3{14}\kappa z_0^2+\fft{129}{2800}\kappa^2z_0^4+\mathcal O(\kappa^3)\right).\numberthis
\end{align*}
Note that the $\mathcal O(\kappa^3)$ and higher terms will depend on the form of the metric functions beyond the order given in (\ref{metricfunctionsorderfourAdS7toAdS4UV}).

\subsubsection{Joining the UV analytic and numeric expressions} 
\label{UVentanglemententropy-c}

It is worth keeping in mind that, so far, we are only considering entangling regions that are fully contained in the UV region of the flow.  In this case, we have used analytic expressions for the metric functions $f_{\mathrm{UV}}(z)$ and $g_{\mathrm{UV}}(z)$.  At the same time, we have expanded the profile of the minimal surface as either $r(z)$ near the $z\to0$ boundary or $z(r)$ near the capoff point $z\to z_0$.  The former expansion allows us to determine the behavior of the UV-divergent terms, (\ref{holographicentanglemententropyAdS7toAdS4UVrofz}), while the latter is a useful starting point for numerical integration.  In particular, the $z(r)$ expansion, (\ref{parameterizingfunctionzUV}), has no free parameters except for the capoff point $z_0$ itself.  This allows for simple numerical integration of $z(r)$ from $r=0$ up to the boundary which is implicitly determined by $z(R)=0$.

The UV minimal surfaces were obtained by numerically solving the $z(r)$ equation, (\ref{differentialequationAdS7toAdS4UVzofr}), transformed into dimensionless variables parametrized by
\begin{align}
    \alpha =\frac{z_0^2\kappa}{4}.
    \label{alpha}
\end{align}
(Note that $\alpha<0$ as we are taking negative curvature $\kappa$.)  We obtain a family of such minimal surfaces parametrized by different values of $\alpha$ using NDSolve in {\it Mathematica} by setting $\textrm{WorkingPrecision} \rightarrow 100$, $\textrm{Method} \rightarrow \textrm{``StiffnessSwitching''}$ and $\textrm{MaxSteps} \rightarrow 10,000$.  Since the $z(r)$ solution becomes singular as the surface approaches $z\to0$, NDSolve will not integrate past this point.  We then invert the numerical $z(r)$ to obtain the profile $r(z)$ near the AdS$_7$ boundary.  This is then fitted to the $z\to0$ solution, (\ref{parameterizingfunctionrUV}), to obtain the radius $R$ of the surface as well as the coefficient $a_{3,0}$ as a function of $\alpha$ (and hence $z_0$).  We call this construction {\it Code 1}.  In addition to obtaining the profile of the UV minimal surfaces, we also integrate (\ref{holographicentanglemententropyAdS7toAdS4preUVzofr}) to obtain the holographic entanglement entropy as a function of $\alpha$.  In this way, we obtain a parametric relation between the entanglement entropy and the radius $R$ of the surface.

We can express the resulting numerically computed holographic entanglement entropy as
\begin{equation}
    S^{\textrm{EE}}_\textrm{UV} \left( R; B^2 \times \mathbb{H}_3, \epsilon \right)
    =
    \frac{2 \pi \mathcal{\ell}^3 \textrm{vol}[\mathbb{H}_3]L_{\mathrm{UV}}^5}{4 G^{(7)}_NR^3}
    \Bigg[
    \frac{1}{4}\left(\fft{R} \epsilon\right)^4
    +
    A \left( \frac{R}{\epsilon} \right)^2
    -
    B\log \left( \epsilon \right)
    +
    C(R)
    +
    \mathcal{O} \left( \epsilon^2 \right)
    \Bigg].
\label{holographicentanglemententropyAdS7toAdS4UV}
\end{equation}
The leading divergent terms were previously calculated in (\ref{holographicentanglemententropyAdS7toAdS4UVrofz}) and whose coefficients $A$ and $B$ are given in (\ref{holographicentanglemententropyAdS7toAdS4UVcoefficientsAB}). Among the divergent terms we verify perfect agreement between the analytical and numerical results.  The term $C(R)$ is a numerical value which depends on $R$ and it is essentially the regulated holographic entanglement entropy. To analyze the behavior of this term we constructed the numerical solutions associated with different values of $R$ through the parameter $\alpha$ defined in (\ref{alpha}). The construction was performed with {\it Code 1} for two different sets of data. For the first set of data we considered  $R \in [0.001, 0.035]$; this is deep in the UV regime where $f(z)$ and $g(z)$ have the same slope (See Fig. \ref{BPSnumericalsolutionAdS7toAdS4version2}). Having obtained precise agreement between the analytic and numerical results in this rage, we consider an extended regime specified by the condition $R \in [0.040, 0.363]$.  This range, while still in the UV, starts departing from the analytic expressions.  The results are presented in figures \ref{SEEreg-a} and \ref{SEEreg-b}, respectively. In Appendix \ref{consistency}, figure \ref{ConsistencySEEreg-a}, we further study the question of matching of the results in the overlapping region.

%%%%%%%%%%%%%%%%%%%%%%%%%%%%%%%%%%%%%%%%%%%%%%%%%%%%%%%%%%%%%%%%%%%%%%%%%%%%%%%%%%%%%%%%%%%%%%%%%%%%%%%%%%%%%%%%%%%%%%%%%%%%%%%%%%%%%%%%%%%%%%%%%%%%%%%%%%%%%
\subsection{IR regime (large entangling region)}
\label{IRentanglemententropy}

We now turn to the IR region, where the geometry is approximated by  AdS$_4\times \mathbb{H}_3$.  As can be seen in Fig.~\ref{BPSnumericalsolutionAdS7toAdS4version2}, this region corresponds to large values of the holographic coordinate $z$, where the function $g(z)$ is approximately constant. Therefore, we declare an entangling region large if its turning point, $z_0$, is in this asymptotic region. In turns of the effective AdS$_4$ radius, $L_{IR}$, this corresponds to $z_0, R\gg L_{\textrm{IR}}$.

The computation of the holographic entanglement entropy (\ref{holographicentanglemententropyAdS7toAdS4}) in the IR regime is a delicate issue due to its sensitivity to the UV part. Technically, the problem is equivalent to the computation of the full entanglement entropy. We will discuss the numerical details of the full computation in Section \ref{wholeentanglemententropy}. Here we focus on a meaningful asymptotic expression for the holographic entanglement entropy (\ref{holographicentanglemententropyAdS7toAdS4}) in the IR regime. Our approximation for the asymptotic expression contains an intermediate cutoff $\Lambda$ that is located already in the region where the background is approximately AdS$_4$. We then consider $z > \Lambda$ as explained in section \ref{BPSIR}. Thus, the asymptotic expression can be estimated as
    \begin{align*}
    S^{\textrm{EE}}_\textrm{IR} \left( R; B^2 \times \mathbb{H}_3, \Lambda \right)
    =
    \frac{2 \pi \mathcal{\ell}^3 \textrm{vol}[\mathbb{H}_3]}{4 G^{(7)}_N}
    \min
    \Bigg[ \displaystyle \int_\Lambda^{z_0} dz \ A_{\textrm{IR}}(z,r(z),r'(z)) \Bigg], \numberthis
    \label{holographicentanglemententropyAdS7toAdS4preIR1}
    \end{align*}
where the Lagrangian $A_{\textrm{IR}}$ is given in the IR region by
    \begin{align}
    A_{\textrm{IR}}(z,r(z),r'(z)) 
    = r(z) \left[ e^{2f_{\textrm{IR}}(z) + 3g_{\textrm{IR}}(z)} \right] \sqrt{1+(r'(z))^2},
    \label{areaAdS7toAdS4IR}
    \end{align}
with the metric functions $f_{\textrm{IR}}$ and $g_{\textrm{IR}}$ given by (\ref{metricfunctionsorderoneAdS7toAdS4IR}). So long as we are sufficiently far in the IR, we can use the leading order terms of the metric functions, which is precisely the IR fixed point solution of the BPS equations, namely
    \begin{equation*}
    f_{\textrm{IR}}(z) \approx \log \left( \frac{L_{\textrm{IR}}}{z} \right), \qquad 
    g_{\textrm{IR}}(z) \approx \frac{1}{2} \log \left( \frac{-\kappa}{2^{1/5} m^2} \right), \numberthis
    \end{equation*}
to rewrite the Lagrangian (\ref{areaAdS7toAdS4IR}) as
    \begin{align*}
    A_{\textrm{IR}}(z,r(z),r'(z)) 
    \approx& \ r(z) \left[ e^{3g_0} \left( \frac{L_{\textrm{IR}}}{z} \right)^2 \right] \sqrt{1+(r'(z))^2} \\
    =& \ e^{3g_0} L_{\textrm{IR}}^2 \ r(z) \frac{\sqrt{1+(r'(z))^2}}{z^2},
    \numberthis
    \label{areaAdS7toAdS4IRleadingorder}
    \end{align*}
where $g_0$ is the IR fixed point value, \ref{coefficientsAdS7toAdS4IR0}). The associated variational principle with respect to the function $r(z)$ leads to the second order non-linear ordinary differential equation
    \begin{align}
    - z \left[ 1+(r'(z))^2 \right] + r(z) \left[ -2 r'(z) -2 (r'(z))^3 + z r''(z) \right]
    = 0.
    \end{align}
The above equation is solved by the semi-circle embedding: 
\begin{align}\label{eq:semi-circle}
    r_{\textrm{IR}}(z) = \sqrt{z_0^2 - z^2}.
    \end{align}
This is the expected result for embedding in AdS spacetimes \cite{Ryu:2006bv}.

One can use the semi-circle embedding, together with the Lagrangian (\ref{areaAdS7toAdS4IRleadingorder}), to rewrite (\ref{holographicentanglemententropyAdS7toAdS4preIR1}) as
    \begin{align*}
    S^{\textrm{EE}}_\textrm{IR} \left( R; B^2 \times \mathbb{H}_3, \Lambda \right)
    \approx&
    \frac{2 \pi \mathcal{\ell}^3 \textrm{vol}[\mathbb{H}_3]}{4 G^{(7)}_N}
    e^{3g_0} L_{\textrm{IR}}^2
    \Bigg[ \displaystyle \int_\Lambda^{z_0} dz \ r_{\textrm{IR}}(z) 
    \frac{\sqrt{1+(r'_{\textrm{IR}}(z))^2}}{z^2} \Bigg] \\
    =&
    \frac{2 \pi \mathcal{\ell}^3 \textrm{vol}[\mathbb{H}_3]}{4 G^{(7)}_N}
    e^{3g_0} L_{\textrm{IR}}^2
    \Bigg[ \displaystyle \int_\Lambda^{z_0} \frac{z_0}{z^2} \ dz
    \Bigg],
    \numberthis
    \label{holographicentanglemententropyAdS7toAdS4preIR2}
    \end{align*}
The holographic entanglement entropy in the IR regime is estimated to be given by
    \begin{align*}
    S^{\textrm{EE}}_\textrm{IR} \left( R; B^2 \times \mathbb{H}_3, \Lambda \right)
    =
    \frac{2 \pi \mathcal{\ell}^3 \textrm{vol}[\mathbb{H}_3]}{4 G^{(7)}_N}
    e^{3g_0} L_{\textrm{IR}}^2
    \Bigg[ \frac{z_0}{\Lambda} - 1
    \Bigg].
    \numberthis
    \label{holographicentanglemententropyAdS7toAdS4IR}
    \end{align*}
If we take $\Lambda$ increasingly small, we see the divergence in (\ref{holographicentanglemententropyAdS7toAdS4IR}) is typical of $\textrm{AdS}_4$. However, as we decrease $\Lambda$ we will eventually leave the regime where the approximation is valid and instead enter an AdS$_7$ regime. The other contribution is universal, that is, $\Lambda$-independent. We will later show that this approximation is vindicated by numerical results.

\begin{figure}[ht]
\centering
    \begin{subfigure}[b]{0.45\textwidth}
    \centering
    \includegraphics[scale=0.78]{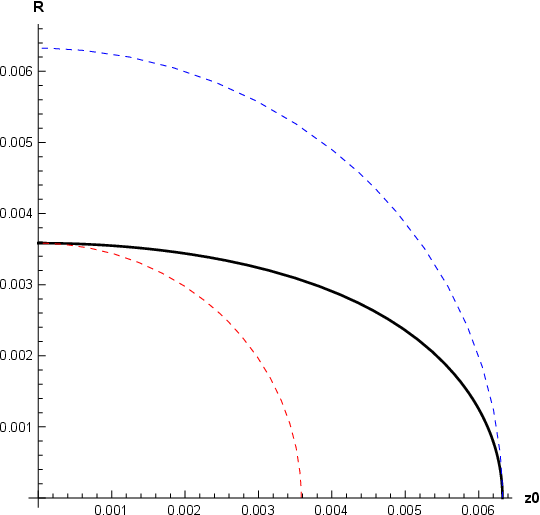}
    \caption{Profile of a minimal surface in the deep UV regime (black line). It is deformed with respect to the semi-circle configurations (red and blue dashed lines).}
    \label{profileUV}
    \end{subfigure}
    \hfill
    \begin{subfigure}[b]{0.45\textwidth}
    \centering
    \includegraphics[scale=0.78]{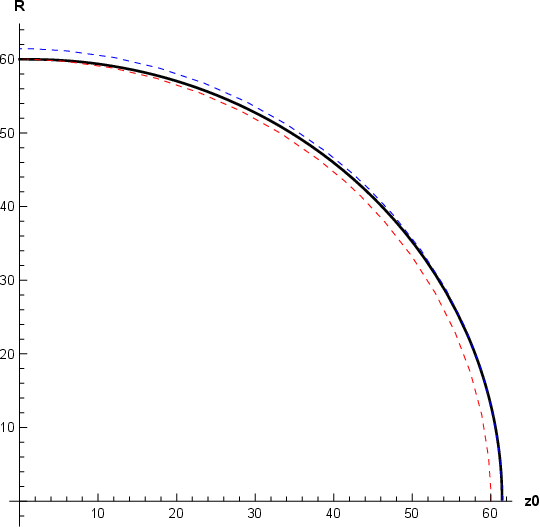}
    \caption{Profile of a minimal surface in the deep IR regime (black line). It is well approximated by the semi-circle configurations (red and blue dashed lines).}
    \label{profileIR}
    \end{subfigure}
\caption{Profiles of the minimal surfaces for both: a) the deep UV regime, and b) the deep IR regime.}
\label{profiles}
\end{figure}

Equation (\ref{eq:semi-circle}) represents the embedding solution corresponding to AdS$_4$. In Fig.~\ref{profiles} we plot the embedding function for numerical solutions corresponding to the regimes of small and large values of $R$, respectively. We have also indicated, using dashed curves, the semi-circle embeddings. It can be seen that for entangling regions remaining mostly in the UV region, corresponding to small values of $R$, the embedding surface is widely deformed with respect to the semi-circle embedding. For large-$R$ entangling surfaces probing the IR, the numerical embedding is well approximated by the semi-circle, lending creed to the approximations following from (\ref{eq:semi-circle}).

%%%%%%%%%%%%%%%%%%%%%%%%%%%%%%%%%%%%%%%%%%%%%%%%%%%%%%%%%%%%%%%%%%%%%%%%%%%%%%%%%%%%%%%%%%%%%%%%%%%%%%%%%%%%%%%%%%%%%%%%%%%%%%%%%%%%%%%%%%%%%%%%%%%%%%%%%%%%%
\subsection{Full entanglement entropy}
\label{wholeentanglemententropy}

\begin{figure}[ht]
\centering
    \begin{subfigure}[b]{0.47\textwidth}
    \centering
    \includegraphics[scale=0.8]{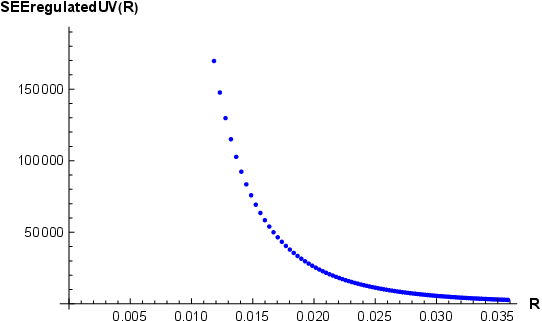}
    \caption{Data for the UV regime defined in the interval $R \in [0.001, 0.035]$.}
    \label{SEEreg-a}
    \end{subfigure}
    \hfill
    \begin{subfigure}[b]{0.47\textwidth}
    \centering
    \includegraphics[scale=0.8]{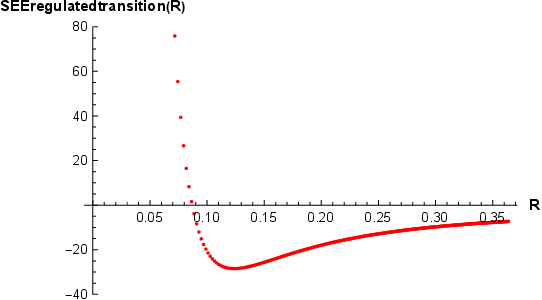}
    \caption{Data for the transition regime defined in the interval $R \in [0.040, 0.363]$.}
    \label{SEEreg-b}
    \end{subfigure} \\ \vspace{0.4cm}
    \begin{subfigure}[b]{0.57\textwidth}
    \centering
    \includegraphics[scale=0.8]{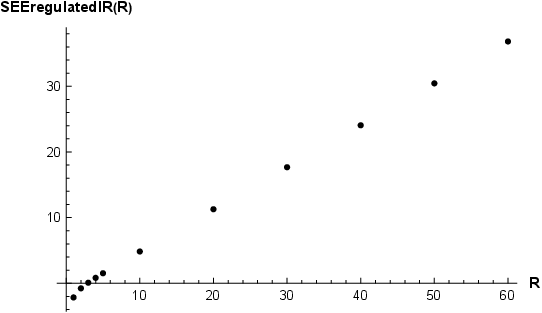}
    \caption{Data for the IR regime defined in the interval $R \in [1, 60]$.}
    \label{SEEreg-c}
    \end{subfigure}
\caption{Regulated holographic entanglement entropy associated with holographic RG flows from $\textrm{AdS}_7$ to $\textrm{AdS}_4 \times \mathbb{H}_3$. The behavior in the IR regime is essentially linear and it decreases in the transition regime till it reaches a minimum before blowing up in the UV regime.}
\label{SEEreg}
\end{figure}

We have so far discussed the behavior of the holographic entanglement entropy both in the UV and the IR regimes in sections \ref{UVentanglemententropy} and \ref{IRentanglemententropy}, respectively. In this section, we present the construction of the regulated holographic entanglement entropy as we move from the UV regime to the IR regime. The construction is associated with both the differential equation (\ref{differentialequationAdS7toAdS4}) and the numerical solution of the BPS equations previously provided in Figure \ref{BPSnumericalsolutionAdS7toAdS4version2}. It is an extension of the regulated holographic entanglement entropy described in section \ref{UVentanglemententropy-c} for larger values of the radius $R$ of the entangling region, and it was performed with NDSolve in {\it Mathematica} by setting $\textrm{WorkingPrecision} \rightarrow 50$, $\textrm{Method} \rightarrow \textrm{``StiffnessSwitching''}$ and $\textrm{MaxSteps} \rightarrow 100,000$. We call this construction {\it Code 2}, and we ran it for a set of sixteen initial values of $R$.  Here both the geometry and the minimal surface was constructed by simultaneous numerical integration from the UV towards the IR.  Once the parameters for the background functions $f(z)$ and $g(z)$ were determined, we then used a shooting method to find the profiles of the minimal surfaces. {\it Code 2} was used to compute the regulated holographic entanglement entropy in the IR regime specified by the condition $R \in [1, 60]$, which can be found in figure \ref{SEEreg-c}. A crucial point here is to verify that the results coming from {\it Code 1} and {\it Code 2} match in the overlapping region. This verification was done in Appendix \ref{consistency}, figure \ref{ConsistencySEEreg-b}. 

The regulated holographic entanglement entropy is obtained by subtracting the UV divergent terms
\begin{equation}
    S^{\textrm{EE}}_\textrm{divergent} \left( R; B^2 \times \mathbb{H}_3, \epsilon \right)
    =
    \frac{2 \pi \mathcal{\ell}^3 \textrm{vol}[\mathbb{H}_3]L_{\mathrm{UV}}^5}{4 G^{(7)}_NR^3}
    \Bigg[
    \frac{1}{4}\left(\fft{R} \epsilon\right)^4
    +
    A \left( \frac{R}{\epsilon} \right)^2
    +
    B\log \left( \fft{L_{\mathrm{UV}}}\epsilon \right)
    \Bigg],
\label{eq:ctsub}
\end{equation}
from the full expression for $S^{\textrm{EE}}$.  This was computed using \textit{Code 1} in the UV and \textit{Code 2} in the IR, and the regulated holographic entanglement entropy is depicted in Figure~\ref{SEEreg}. The data of Figure~\ref{SEEreg} can be used to estimate the regulated holographic entanglement entropy in the UV and IR regimes, respectively, as
    \begin{align}
    S^{\textrm{UV}}_{\textrm{EEreg}} \approx -0.34293 \frac{1}{R^3} -0.14059 \frac{1}{R^3} \log(R),
    \label{regulatedholographicentanglemententropyAdS7toAdS4UVnumerical}
    \end{align}
and
    \begin{align}
    S^{\textrm{IR}}_{\textrm{EEreg}} \approx -1.21611 \frac{1}{R} + 0.63765 R -1.41417.
    \label{regulatedholographicentanglemententropyAdS7toAdS4IRnumerical}
    \end{align}
A few important checks of these numerical results are in order.  Firstly, we note the presence of a $(1/R^3)\log(R)$ term in the UV.  Moreover, its coefficient matches precisely the first term of coefficient $B$ in (\ref{holographicentanglemententropyAdS7toAdS4UVrofz}).  This term naturally arises in the UV, but has not been subtracted out since we used only $\log(1/\epsilon)$ and not $\log(R/\epsilon)$ in the counterterm subtraction of (\ref{eq:ctsub}).  We will return to this point below in Section~\ref{Thecfunction}.  Secondly, as can be seen in Figure~\ref{SEEreg-c}, $S^{\textrm{IR}}_{\textrm{EEreg}}$ asymptotically approaches a linear function of $R$ in the IR.  This matches the analytic behavior in (\ref{holographicentanglemententropyAdS7toAdS4IR}) where $z_0\approx R$ since the minimal surface is well approximated by the semi-circle in the deep IR.  Note that the constant term in  (\ref{holographicentanglemententropyAdS7toAdS4IR}) is universal, and matches the constant term in the numerical expression,(\ref{regulatedholographicentanglemententropyAdS7toAdS4IRnumerical}).

We can immediately observe from Figure~\ref{SEEreg} that the regulated holographic entanglement entropy is \textit{not} monotonic with the radius $R$ of the entangling surface.  In particular, it decreases away from the UV, reaches a minimum in the transition region, and then increases linearly in the IR.  We will revisit this behavior in Section~\ref{Thecfunction} when constructing a $c$-function.  But, first, we proceed with a general discussion of the particular results above from a field-theoretic and wider holographic point of view.

%%%%%%%%%%%%%%%%%%%%%%%%%%%%%%%%%%%%%%%%%
\section{The \texorpdfstring{$c$}{c}-function}
\label{Thecfunction}

We have investigated the holographic entanglement entropy for a flow from AdS$_7$ to AdS$_4\times\mathbb H_3$, both analytically in the UV and IR in sections \ref{UVentanglemententropy} and \ref{IRentanglemententropy}, respectively, and numerically throughout the flow in section \ref{wholeentanglemententropy}.  Given this entanglement entropy, it is natural to explore the possibility of constructing a $c$-function for this flow across dimensions.

We start by considering the entropic $c$-function \cite{Casini:2017vbe}
\begin{equation}
    c(R)=(R\partial_R-(d-2))\Delta S(R),
\end{equation}
where $d$ is the dimension of the field theory and $\Delta S=S_{\textrm{deformed}}-S_{\textrm{CFT}}$.  It was shown in \cite{Casini:2017vbe}, using strong sub-additivity, that this $c$-function is monotonic along the flow, namely $c'(R)\le0$.  In order to apply this expression to the holographic context for flows across dimension, we must consider both the holographic version of $\Delta S$ and the dimension $d$.  For the latter, since we are considering a flow to AdS$_4$/CFT$_3$, it is natural to take $d=3$ (\textit{i.e.} the dimension of the IR theory).  At the same time, the subtraction of $S_{\textrm{CFT}}$  in $\Delta S$ is intended to regulate the UV.  Thus it may seem natural to choose $S_{\textrm{CFT}}$ to be the holographic entanglement entropy of a spherical entangling surface in AdS$_7$ \cite{Ryu:2006bv,Ryu:2006ef}
\begin{equation}
    S_{\textrm{EE}}^{\textrm{CFT}}(R;B^5,\epsilon)=\fft{8\pi^2}3\fft{L_{\mathrm{UV}}^5}{4G_N^{(7)}}\left[\fft14\left(\fft{R}\epsilon\right)^4-\fft34\left(\fft{R}\epsilon\right)^2+\fft38\log\left(\fft{2R}\epsilon\right)+\fft9{32}\right].
\end{equation}
However, it is important to realize that, even in the UV, the minimal surface wraps $\mathbb H_3$ in AdS$_7$.  Hence what we really want is to take a $B^2\times\mathbb H_3$ surface.  Even with this in mind, we note that the negative curvature of $\mathbb H_3$ necessarily induces a flow away from the UV fixed point.  Thus it is not obvious if there is even a well defined $S_{\textrm{CFT}}$ that can be used for the subtraction.

One possibility for the UV subtraction is to take $\kappa=0$ for the internal 3-manifold.  This ensures that there is an AdS$_7$ fixed point with%
\footnote{These coefficients can be obtained by setting $\kappa=0$ in (\ref{holographicentanglemententropyAdS7toAdS4UVrofz}).}
\begin{equation}
    S_{\textrm{EE}}^{\textrm{CFT}}(R;B^2\times M_3,\epsilon)=\fft{2\pi\ell^3\mathrm{vol}[M_3]L_{\mathrm{UV}}^5}{4G_N^{(7)}R^3}\left[\fft14\left(\fft{R}\epsilon\right)^4-\fft3{64}\left(\fft{R}\epsilon\right)^2-\fft9{2048}\log\left(\fft{R}\epsilon\right)+\mathcal O(\epsilon^0)\right].
\end{equation}
At the same time, however, this is not asymptotic to the same conformal boundary as the flowing theory, so it does not cancel all the divergences as highlighted in (\ref{eq:ctsub}).  Nevertheless, this motivates us to define the subtracted entanglement entropy $\Delta S=S_{\mathrm{deformed}}-\bar S$ where
\begin{align}
    \bar S&=\fft{2\pi\ell^3\mathrm{vol}[M_3]L_{\mathrm{UV}}^5}{4G_N^{(7)}R^3}\biggl[\fft14\left(\fft{R}\epsilon\right)^4-\fft3{64}\left(\fft{R}\epsilon\right)^2-\fft9{2048}\log\left(\fft{R}\epsilon\right)\nn\\
   &\kern12.2em -\fft{\kappa R^2}8\left(\fft{R}\epsilon\right)^2+\fft{3(5\kappa R^2+2\kappa^2R^4)}{640}\log\left(\fft{L_{\mathrm{UV}}}\epsilon\right)\biggr]\nn\\
   &=\fft{2\pi\ell^3\mathrm{vol}[\mathbb H_3]L_{\mathrm{UV}}^5}{4G_N^{(7)}R^3}\biggl[\fft14\left(\fft{R}\epsilon\right)^4+A\left(\fft{R}\epsilon\right)^2+B\log\left(\fft{L_{\mathrm{UV}}}\epsilon\right)-\fft9{2048}\log\left(\fft{R}{L_{\mathrm{UV}}}\right)\biggr].
\end{align}
Here, $\bar S$ is the $\kappa=0$ entanglement entropy supplemented by $\kappa R^2$ dependent terms that are necessary to remove the curvature-induced divergences.  Note that $\bar S$ differs from $S_{\mathrm{divergent}}^{\mathrm{EE}}$ given in (\ref{eq:ctsub}) by the last term in the second line, namely the term proportional to $(1/R^3)\log(R)$.  This has the effect of removing the log term in the regulated UV entanglement entropy, (\ref{regulatedholographicentanglemententropyAdS7toAdS4UVnumerical}), but does not modify the IR behavior of (\ref{regulatedholographicentanglemententropyAdS7toAdS4IRnumerical})%
\footnote{Note that simply modifying (\ref{eq:ctsub}) by replacing $\log(L_{\mathrm{UV}}/\epsilon)$ by $\log(R/\epsilon)$ will modify the IR behavior by introducing a term proportional to $\kappa^2 R\log(R)$.  This IR modification is undesirable, as the $\bar S$ subtraction is intended to address UV sensitive effects only.}

\begin{figure}[ht]
\centering
    \begin{subfigure}[b]{0.47\textwidth}
    \centering
    \includegraphics[scale=0.8]{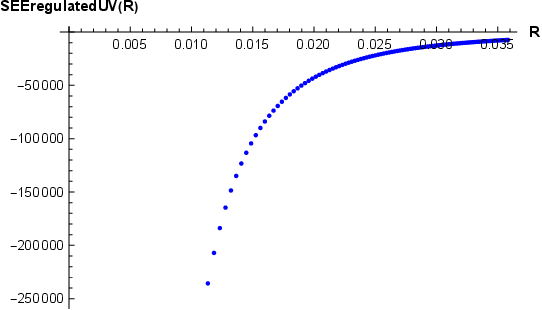}
    \caption{Data for the UV regime defined in the interval $R \in [0.001, 0.035]$.}
    \label{SEEreg2-a}
    \end{subfigure}
    \hfill
    \begin{subfigure}[b]{0.47\textwidth}
    \centering
    \includegraphics[scale=0.8]{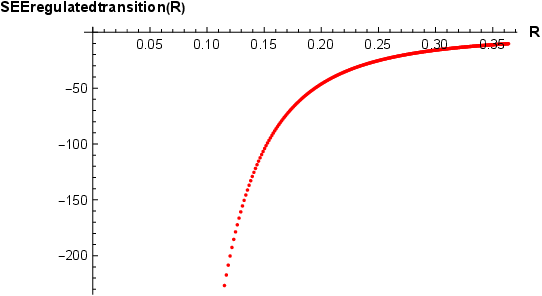}
    \caption{Data for the transition regime defined in the interval $R \in [0.040, 0.363]$.}
    \label{SEEreg2-b}
    \end{subfigure} \\ \vspace{0.4cm}
    \begin{subfigure}[b]{0.57\textwidth}
    \centering
    \includegraphics[scale=0.8]{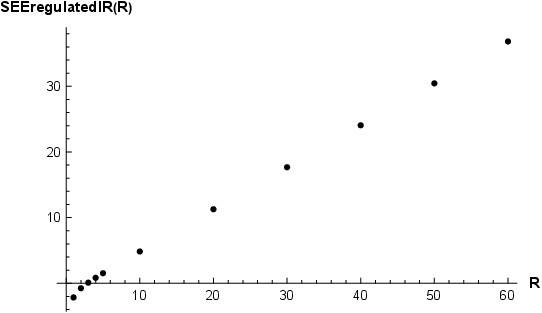}
    \caption{Data for the IR regime defined in the interval $R \in [1, 60]$.}
    \label{SEEreg2-c}
    \end{subfigure}
\caption{Subtracted holographic entanglement entropy, (\ref{eq:SEEsubtracted}), for holographic RG flows from $\textrm{AdS}_7$ to $\textrm{AdS}_4 \times \mathbb{H}_3$.}
\label{SEEreg2}
\end{figure}

The subtracted entanglement entropy
\begin{equation}
    \Delta S_{\mathrm{EE}}=S_{\mathrm{EE}}(R;B^2\times\mathbb H_3,\epsilon)-\bar S,
\label{eq:SEEsubtracted}
\end{equation}
is depicted in figure \ref{SEEreg2}. It can be observed that $\Delta S$ is has a monotonically increasing behavior.  Moreover, the IR regime is linearly growing with the same asymptotic behavior as before in (\ref{regulatedholographicentanglemententropyAdS7toAdS4IRnumerical}).  To demonstrate consistency, we had addressed the matching of the data in the overlapping regions, and the relevant plots can be seen in Appendix \ref{consistency}, Fig.~\ref{ConsistencySEEreg2}.  An approximate fit to the data in the various regions gives
\begin{subequations}
    \begin{align*}
    \Delta S^{\textrm{UV}}_{\textrm{EE}} & \approx -0.34313 \frac{1}{R^3}, \numberthis
    \label{SEEreg2fit-a} \\ \\
    \Delta S^{\textrm{transition}}_{\textrm{EE}} & \approx -0.34380 \frac{1}{R^3} + 0.05486 \frac{1}{R^2} -0.25261 \frac{1}{R} + 3.39700 R -4.06244, \numberthis
    \label{SEEreg2fit-b} \\ \\
    \Delta S^{\textrm{IR}}_{\textrm{EE}} & \approx  -1.22289 \frac{1}{R} + 0.63764 R -1.41390,
    % -0.09568 \frac{1}{R^4} -0.14013 \frac{1}{R^3} + 0.08076 \frac{1}{R^2}
    \numberthis
    \label{SEEreg2fit-c}
    \end{align*}
    \label{SEEreg2fit}%
\end{subequations}
As noted, $\Delta_{\mathrm{EE}}^{\mathrm{UV}}$ in (\ref{SEEreg2fit-a}) is consistent with the $1/R^3$ term in the regulated UV expression, (\ref{regulatedholographicentanglemententropyAdS7toAdS4UVnumerical}). Analogously, $\Delta_{\mathrm{EE}}^{\mathrm{IR}}$ in  (\ref{SEEreg2fit-c}) remains approximately the same as the regulated IR expression, (\ref{regulatedholographicentanglemententropyAdS7toAdS4IRnumerical}).

\begin{figure}[ht]
\centering
    \begin{subfigure}[b]{0.47\textwidth}
    \centering
    \includegraphics[scale=0.8]{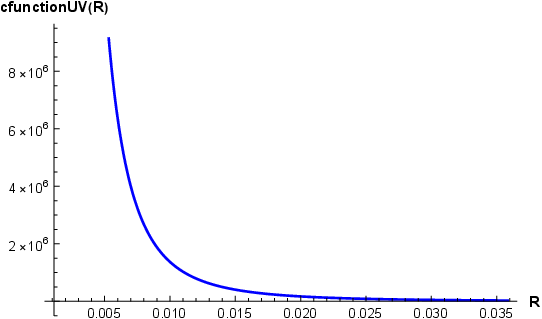}
    \caption{UV regime $c$-function defined in the interval $R \in [0.001, 0.035]$.}
    \label{cfunction2-a}
    \end{subfigure}
    \hfill
    \begin{subfigure}[b]{0.47\textwidth}
    \centering
    \includegraphics[scale=0.8]{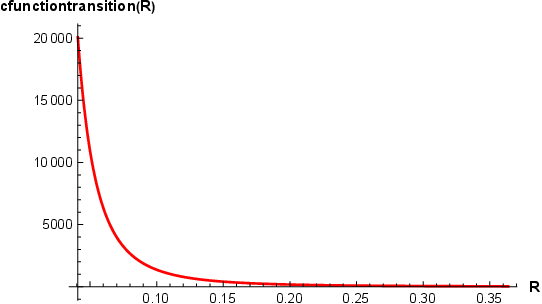}
    \caption{Transition regime $c$-function defined in the interval $R \in [0.040, 0.363]$.}
    \label{cfunction2-b}
    \end{subfigure} \\ \vspace{0.4cm}
    \begin{subfigure}[b]{0.47\textwidth}
    \centering
    \includegraphics[scale=0.8]{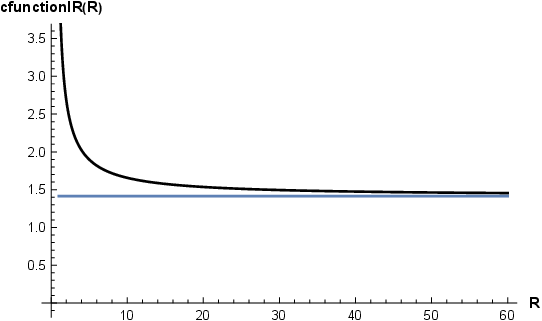}
    \caption{IR regime $c$-function (black line) defined in the interval $R \in [1, 60]$. It approaches the value of the IR central charge (blue line) as $R$ increases.}
    \label{cfunction2-c}
    \end{subfigure}
\caption{Monotonically decreasing non-interpolating $c$-function for holographic RG flows from $\textrm{AdS}_7$ to $\textrm{AdS}_4 \times \mathbb{H}_3$ as defined in (\ref{eq:holoregc}).  The $c$-function is displayed in the UV, transition and IR regions.}
\label{cfunction2}
\end{figure}

We now define the holographic regulated $c$-function \cite{Casini:2017vbe}
\begin{equation}
    c(R)=(R\partial_R-1)\Delta S_{\mathrm{EE}}(R).
\label{eq:holoregc}
\end{equation}
Since $\Delta S_{\mathrm{EE}}$ has been determined numerically, instead of taking a numerical derivative to obtain the $c$-function, we work with the approximate analytical expressions, (\ref{SEEreg2fit}).  It is verified in Appendix~\ref{consistency}, Fig.~\ref{Consistencycfunction2}, that the $c$-function obtained from the numerical data matches in the overlapping regions.  In the UV, $c(R)$ falls off as $c^{\mathrm{UV}}\approx1.373/R^3$, while it levels off and reaches a constant value $c^{\mathrm{IR}}\approx1.414$ in the IR.  In particular, this function is monotonically decreasing along the flow, as depicted in Figure~\ref{cfunction2}.  Since this $c$-function effectively counts CFT$_3$ degrees of freedom, it blows up in the UV, but flows to the IR central charge, which is given analytically as $c^{\mathrm{IR}}=\sqrt2$.  Consequently, we have constructed a monotonically decreasing non-interpolating $c$-function.

It remains to be seen whether there is a natural interpolating $c$-function that connects the CFT$_6$ central charge in the UV to the CFT$_3$ central charge in the IR.  Of course, if one were to remain in the UV regime, the corresponding central charge was discussed and placed in a field-theoretic framework in Section~\ref{sec:anomalycoefs}.

%%%%%%%%%%%%%%%%%%%%%%%%%%%%%%%%%%%%%%%%%%%%%%%%%%%%%%%%%%%
\section{Anomaly coefficients and entanglement entropy}
\label{sec:anomalycoefs}

While we have mainly focused on a holographic RG flow, ultimately one would like to make a connection to the dual field theory.  We thus consider a field theory defined on a $d$-dimensional manifold $M$ foliated by $(d-1)$-dimensional spatial slices.  To proceed, we divide the space into two regions, $A$ and its complement $\bar A$, separated by the entangling surface $\Sigma$.  The entanglement entropy is then given by the von~Neumann entropy of the reduced density matrix that is obtained by tracing over $\bar A$.  Note that $\Sigma$ is taken as a constant time surface so it is codimension-two in $M$.

The entanglement entropy in a field theory is UV divergent, and the leading divergence scales as the area of $\Sigma$.  In addition, there are subleading divergences, so that
\begin{equation}
    S_{\mathrm{EE}}=\gamma_1\fft{\mbox{Area}(\Sigma)}{\epsilon^{d-2}}+\fft{a_2}{\epsilon^{d-4}}+\fft{a_3}{\epsilon^{d-6}}+\cdots,
\end{equation}
where $\epsilon$ is a UV cutoff with dimension length.  Furthermore, for even-dimensional theories, this set of divergences ends in a log divergence of the form $a_{d/2}\log\epsilon$.  The coefficient of this log term is particularly interesting as it has been shown to exhibit universal behavior across a whole range of theories.

The universality of the log divergence is especially pronounced for conformal field theories where it can be related to the anomaly coefficients of the theory.  For a two-dimensional CFT, we have
\begin{equation}
    S_{\mathrm{EE}}^{d=2}=\fft{c}3\log\fft\ell\epsilon+\cdots,
\end{equation}
where $\ell$ is the length of the interval $A$ and $c$ is the two-dimensional central charge.  On the other hand, four dimensional CFTs are characterized by two central charges, $c$ and $a$, and the coefficient of the log term depends not just on these anomaly coefficients but also on the geometry of the entangling surface $\Sigma$ and how it is embedded in $M$.  A combination of field theoretic and holographic computations gives the expression
\begin{equation}
    S_{\mathrm{EE}}^{d=4}=\gamma_1\fft{\mbox{Area}(\sigma)}{\epsilon^2}+a_2\log\fft\ell\epsilon+\cdots,
\end{equation}
where \cite{Ryu:2006ef,Solodukhin:2008dh,Hung:2011xb}
\begin{equation}
    a_2=\fft1{2\pi}\int_\Sigma d^2x\sqrt{h}\left[c\left(C^{abcd}h_{ac}h_{bd}-K_a^{\hat i\,b}K_b^{\hat i\, a}+\fft12K_a^{\hat i\,a}K_b^{\hat i\,b}\right)-a\mathcal R\right].
\label{eq:a2coef}
\end{equation}
Here $C_{abcd}$ is the four-dimensional Weyl tensor on $M$, $h_{ab}$ is the pullback of the metric onto $\Sigma$, and $\mathcal R$ is the two-dimensional Ricci scalar on $\Sigma$ (\textit{i.e.}\ the intrinsic curvature).  Since $\Sigma$ is codimension two, there are two surface normals, $n^{\hat i}$ with $\hat i=1,2$, with corresponding extrinsic curvature tensors $K^{\hat i}_{ab}$.

Since the coefficient of the log term depends on both the background $M$ and the entangling surface $\Sigma$, we can probe the anomaly coefficients, $c$ and $a$, by varying a combination of $M$ and $\Sigma$.  For a conformally flat background, the Weyl tensor vanishes, and the $a_2$ coefficient depends only on the intrinsic and extrinsic curvatures of $\Sigma$.  If, in addition, $\Sigma$ is a minimal surface, then its extrinsic curvature vanishes, and the log coefficient is only sensitive to the $a$ central charge.  However, more generally, all terms in (\ref{eq:a2coef}) may contribute.

As an example of how one may extract the central charges $c$ and $a$ by using different entangling surfaces, $\Sigma$, consider a CFT on a flat background, $M=\mathbb R^{1,3}$.  For $\Sigma=S^2$ a sphere of radius $R$, one finds $a_2(S^2)=-4a$, while for $\Sigma=S^1\times\mathrm R$ a right-circular cylinder with circle radius $R$, one finds instead $a_2(S^1\times\mathrm R)=-(\ell/2R)c$, where the length of the cylinder, $\ell$, is introduced as a cutoff.

%%%%%%%%%%%%%%%%%%%%%%%%%%%%%%%%%%%%%%%
\subsection{The log coefficient for six-dimensional CFTs}

As we are considering a holographic flow from an AdS$_7$ UV theory, our main interest is in the universal coefficient of the logarithmic term of the entanglement entropy associated with the four-dimensional entangling surface $\Sigma$ in CFT defined on a six-dimensional background $M$.  Here, the situation is more involved, as the six-dimensional trace anomaly involves four central charges.  In particular, following the notation of \cite{Hung:2011xb}, one has
\begin{equation}
    \langle T^i_i\rangle=\sum_{n=1}^3B_nI_n+2AE_6,
\end{equation}
where $I_n$ are built from the Weyl tensor
\begin{equation}
    I_1=C_{kijl}C^{imnj}C_m{}^{kl}{}_n,\qquad I_2=C_{ij}{}^{kl}C_{kl}{}^{mn}C_{mn}{}^{ij},\qquad I_3=C_{iklm}(\Box\delta^i_j+4R^i_j-\ft65R\delta^i_j)C^{jklm},
\end{equation}
and $E_6$ is the six-dimensional Euler density.

In general, the divergent terms in the six-dimensional entanglement entropy take the form
\begin{equation}
    S_{EE}^{d=6}=\gamma_1\fft{\mathrm{Area}(\Sigma)}{\epsilon^4}+\fft{a_2}{\epsilon^2}+a_3\log\fft\ell\epsilon+\cdots,
\end{equation}
and again we are interested in the coefficient of the log term, which in this case is denoted by $a_3$.  On general grounds, one expects it to take the schematic form
\begin{equation}
    a_3\sim\int_\Sigma d^4x\sqrt{h}\left[\sum_{n=1}^3B_n\left(C^2+CK^2+K^4\right)_n+AE_4\right],
\label{eq:a3schem}
\end{equation}
where $C$ denotes the Weyl tensor, $K$ the extrinsic curvature of $\Sigma$, and $E_4$ is the four-dimensional Euler density on $\Sigma$. It is plausible that $a_3$ above contains the necessary ingredients toward a definition of $c(\widetilde{\rm UV})$ in Section \ref{HolographicEE}; we will pursue this direction elsewhere.

The precise expression for $a_3$ is quite involved, and was constructed in reference \cite{Hung:2011xb} for the case of zero extrinsic curvature, with the result
\begin{align}
a_3\Big|_{K_{ab}^{\hat i}=0} = \int d^4 x \sqrt{h}\left[ 2 \pi \sum_{n=1}^{3} B_n
\frac{\partial I_n}{\partial R^{ij}_{\ \ kl}} \tilde{\varepsilon}^{ij} \tilde{\varepsilon}_{kl} + 2 A E_4
\right]_{\Sigma},
\label{coefficientMyers}
\end{align}
with
\begin{align*}
\frac{\partial I_1}{\partial R^{ij}_{\ \ kl}} \tilde{\varepsilon}^{ij} \tilde{\varepsilon}_{kl} &=
3 \left( C^{jmnk} C^{\ \ il}_{m \ \ n} \tilde{\varepsilon}_{ij} \tilde{\varepsilon}_{kl} 
- \frac{1}{4} C^{iklm} C^{j}_{\ klm} \tilde{g}^{\perp}_{ij} 
+ \frac{1}{20} C^{ijkl} C_{ijkl} \right), \\
\frac{\partial I_2}{\partial R^{ij}_{\ \ kl}} \tilde{\varepsilon}^{ij} \tilde{\varepsilon}_{kl} &=
3 \left( C^{klmn} C_{mn}^{\ \ \ ij} \tilde{\varepsilon}_{ij} \tilde{\varepsilon}_{kl} 
- C^{iklm} C^{j}_{\ klm} \tilde{g}^{\perp}_{ij} 
+ \frac{1}{5} C^{ijkl} C_{ijkl} \right), \\
\frac{\partial I_3}{\partial R^{ij}_{\ \ kl}} \tilde{\varepsilon}^{ij} \tilde{\varepsilon}_{kl} &=
2 \left( \Box C^{ijkl} + 4 R^{i}_{\ m} C^{mjkl} - \frac{6}{5} R C^{ijkl} \right) \tilde{\varepsilon}_{ij} \tilde{\varepsilon}_{kl} \\
& \ \ \ - 4 C^{ijkl} R_{ik} \tilde{g}^{\perp}_{jl} + 4 C^{iklm} C^{j}_{\ klm} \tilde{g}^{\perp}_{ij} 
- \frac{12}{5} C^{ijkl} C_{ijkl}.
\numberthis
\label{BtermsMyers}
\end{align*}
In the above expressions, $R_{ijkl}$, $C_{ijkl}$, $R_{ij}$ and $R$ stand for the $6$-dimensional Riemann tensor, Weyl tensor, Ricci tensor and Ricci scalar, respectively, while $\tilde{\varepsilon}_{ij}$ and $\tilde{g}^{\perp}_{ij}$ stand for the Levi-Civita symbol and the metric of the geometry $M \setminus \Sigma$. Furthermore, $h$ is the metric of $M$ induced on $\Sigma$ and $E_4$ is the $4$-dimensional Euler density
\begin{align}
E_4 = R^{ijkl} R_{ijkl} - 4 R^{ij} R_{ij} + R^2.
\label{Euler4Myers}
\end{align}

While the zero extrinsic curvature case is sufficient when $\Sigma$ is a minimal surface, many entangling surfaces will have extrinsic curvature, in which case the additional terms related to $K$ in (\ref{eq:a3schem}) will contribute.  This is in fact the situation we are faced with in the analysis of the holographic RG flow from $\textrm{AdS}_7$ to $\textrm{AdS}_4 \times \mathbb{H}_3$.  Here, the background field theory geometry is $\mathbb{R}^3 \times \mathbb{H}_3$ and the associated entangling surface is $S^1 \times \mathbb{H}_3$ with $S^1$ having non-vanishing extrinsic curvature in $\mathbb R^3$.  The divergent terms in the holographic entanglement entropy are given in (\ref{holographicentanglemententropyAdS7toAdS4UVrofz}) with coefficients given in (\ref{holographicentanglemententropyAdS7toAdS4UVcoefficientsAB}).  Focusing on the log term, we find, in particular
\begin{equation}
    a_3=\mathrm{Area}(\Sigma)\fft{L_{\mathrm{UV}}^5}{4G_N^{^(7)}}\left[\fft3{80}\left(\fft1{\ell^2}\right)^2-\fft3{64}\left(\fft1{\ell^2}\right)\left(\fft1{R^2}\right)-\fft9{2048}\left(\fft1{R^2}\right)^2\right],
\label{eq:logcoef}
\end{equation}
where $\mathrm{Area}(\Sigma)=(2\pi R)(\ell^3\mathrm{vol}[\mathbb H_3])$, and where we have furthermore restored the radius of curvature of $\mathbb H_3$ by taking $\kappa=-2/\ell^2$.  This expression for $a_3$ has the structure of (\ref{eq:a3schem}), where the Weyl tensor on $M$ measures the curvature of $\mathbb H_3$, so that $C_{ijkl}\sim1/\ell^2$, and the extrinsic curvature of $\Sigma$ is that of the $S^1$ embedded in $\mathbb R^3$, so that $K_{ab}^{\hat i}\sim1/R$.  (Note that the Euler characteristic vanishes for $\Sigma=S^1\times\mathbb H_3$, so this surface is not sensitive to the $A$ anomaly.)

It is important to keep in mind that the log coefficient we obtained in (\ref{eq:logcoef}) is for a flowing theory, while the field theory expressions, (\ref{eq:a3schem}) and (\ref{coefficientMyers}), are for a strictly conformal theory.  Nevertheless, it is instructive to compare the holographic and CFT expressions.  Taking $M=\mathbb R^3\times\mathbb H_3$ and $\Sigma=S^1\times\mathbb H_3$, the expression (\ref{coefficientMyers}) gives
\begin{equation}
    a_3\Big|_{K^{\hat i}_{ab}=0}=- \frac{9\pi}{25}\textrm{Area}(\Sigma)\fft{9 B_1 - 4 B_2 + 64 B_3}{\ell^4},
\label{termKappa0}
\end{equation}
where $\ell$ is the `radius' of $\mathbb H_3$.  To compare with the gravity result, we may use the holographic Weyl anomaly dictionary \cite{Henningson:1998gx}
\begin{equation}
    B_1=4B_2=-12B_3=-\fft{A}{16\pi^3}=-\fft1{32}\fft{L_{\mathrm{UV}}^5}{2\pi\cdot4 G_N^{(7)}},
\end{equation}
to obtain
\begin{equation}
    a_3=\mathrm{Area}(\Sigma)\fft{L_{\mathrm{UV}}^5}{4G_N^{(7)}}\left[\fft3{200\ell^4}+\fft{\alpha_1}{\ell^2R^2}+\fft{\alpha_2}{R^4}\right].
\label{termKappa0reduced}
\end{equation}
Here we have parametrized the unknown extrinsic curvature contributions by the constants $\alpha_1$ and $\alpha_2$ in order to highlight the dependence on the curvature of $M$ ($C_{ijkl}\sim 1/\ell^2$) and the extrinsic curvature of $\Sigma$ ($K_{ab}^{\hat i}\sim 1/R$).

One immediately notices that the coefficient of the $(\mathrm{Weyl})^2$ terms in (\ref{eq:logcoef}) and (\ref{termKappa0reduced}) do not agree, even though the two expressions have a similar dependence on $\ell$ and $R$.  While this at first may come as a surprise, the resolution is that the log coefficient, while universal in a CFT, can in fact pick up additional contributions in a flowing theory, as discussed in \cite{Hung:2011ta}.  To see how this occurs, we now turn to the holographic theory.

%%%%%%%%%%%%%%%%%%%%%%%%%%%%%%%%%%%%%%%%%%%%%%%%%%%%%%%%%%%%%%%%%%%%%%%%%%%%%%%%%%%%%%%%%%%%%%%%%%%%%%%%%%%%%%%%%%%%%%%%%%%%%%%%%%%%%%%%%%%%%%%%%%%%%%%%%%%%%
\subsection{Entanglement entropy from the holographic dual}

In the asymptotically AdS bulk, the divergent contribution to the holographic entanglement entropy occurs at the boundary.  Therefore, it is sufficient to expand near the boundary in order to identify the divergence.  To highlight the UV behavior, we find it convenient to pull out the asymptotic AdS factor from the metric (\ref{metricDd}).  We thus take
\begin{align}
	ds^2 = \left( \frac{L_{\textrm{UV}}}{z}\right)^2 \left[ e^{2\hat{f}(z)} \left( -dt^2 + dz^2 + dr^2 + r^2 d\Omega^2_{d-2} \right) + e^{2\hat{g}(z)} \mathcal{\ell}^{D-d} ds^2_{M_{D-d}} \right],
    \label{metricDdcomparison1}
\end{align}
where the metric functions $\hat f$ and $\hat g$ approach zero as $z\to0$ in the UV.  The expression for the holographic entanglement entropy, (\ref{holographicentanglemententropyDd}), then becomes
\begin{align}
    & S_{\textrm{EE}} \left( R; B^{d-1} \times M_{D-d}, \epsilon \right) \\
    =&
    \frac{\textrm{vol}[S^{d-2}] \ \mathcal{\ell}^{D-d} \ \textrm{vol}[M_{D-d}]}{4 G^{(D+1)}_N}
    \min_{r(\epsilon)=R}
    \Bigg[ \displaystyle \int_\epsilon^{z_0} dz \left( \frac{L_{\textrm{UV}}}{z}\right)^{D-1} (r(z))^{d-2} \ e^{(d-1)\tilde{F}(z)}
    \sqrt{1+(r'(z))^2} \Bigg], \numberthis
\label{holographicentanglemententropyDdcomparison1}
\end{align}
where the warp factor $\tilde{F}$ is defined in a similar manner as (\ref{warpfactorDd}), except with the subtracted functions $\hat f$ and $\hat g$
\begin{align}
	\tilde{F}(z) = \hat{f}(z) + \frac{D-d}{d-1}\hat{g}(z).
	\label{warpfactorDdcomparison1}
\end{align}

We now specialize to the case of flows from AdS$_7$ to AdS$_4\times M_3$, corresponding to $D=6$ and $d=3$.  The minimal surface in (\ref{holographicentanglemententropyDdcomparison1}) depends only on the bulk geometry through the function $\tilde F$.  Motivated by the Fefferman-Graham expansion, we assume the UV behavior
\begin{align}
    \tilde{F}(z) = \tilde f_1 z^2 + \tilde f_2 z^4 + \mathcal{O} \left(\{z^6,z^6\log z\} \right),
\label{Fhatexpansioncomparison1}
\end{align}
for some constants $\tilde f_1$ and $\tilde f_2$.  In the absence of bulk matter, this is an expansion in even powers of $z$ along with higher and higher powers of $\log z$ the further one expands.  More generally, this may need to be modified if there is bulk matter triggering a relevant deformation, as the backreaction of the matter on the metric will introduce terms of the form $z^{2\Delta}$ along with higher powers where the conformal weight $\Delta$ need not be an integer.  Thus the general expansion of $\tilde F(z)$ can be rather complicated.  But, as shown in \cite{Hung:2011ta}, these backreaction terms will not contribute to the log divergence, \textit{unless} they appear at integer powers of $z$, in which case an expansion of the form (\ref{Fhatexpansioncomparison1}) remains valid.  This is, in fact, the case for the AdS$_7$ flow, as one can see from the UV regime expansion, (\ref{metricfunctionsorderfourAdS7toAdS4UV}).

Given the metric parametrization, (\ref{Fhatexpansioncomparison1}), we can solve for the minimum surface asymptotically near the boundary.  The result is
\begin{equation}
    r(z)=R\left(1-\fft18\left(\fft{z}R\right)^2-\fft{7+64\tilde f_1R^2}{512}\left(\fft{z}R\right)^4+\cdots\right),
\end{equation}
which leads to the asymptotic behavior of the holographic entanglement entropy
\begin{align}
    S_{\textrm{EE}}\left( R; B^2 \times M_3, \epsilon \right)
    &=
    \frac{(2\pi R)\mathcal{\ell}^{3} \textrm{vol}[M_3] L_{\textrm{UV}}^5}{4 G^{(7)}_N}
    \Bigg[\frac{1}{4\epsilon^4} + \left( \tilde f_1-\frac3{64 R^2} \right) \frac{1}{\epsilon^2} \nn\\
    &\kern1cm +\left(2 \left(\tilde f_1^2 + \tilde f_2 \right) -\frac{ 3\tilde f_1}{16 R^2} - \frac9{2048 R^4} \right) \log\left( \frac{R}{\epsilon} \right) +\mbox{finite}\Bigg],
\label{holographicentanglemententropyAdS7toAdS4UVrofzcomparison1}
\end{align}
where the dependence on the metric arises through the coefficients $\tilde f_1$ and $\tilde f_2$.

Since the minimal surface wraps the entire $M_3$, the entanglement entropy is not directly sensitive to its geometry, but only to its volume.  It is important to note, however, that the geometry of $M_3$ enters indirectly through the bulk Einstein equation determining $\tilde f_1$ and $\tilde f_2$.  As an example, and to make connection with the CFT result, (\ref{termKappa0reduced}), we can take the internal manifold to be $\mathbb{H}_3$ and assume the vacuum AdS Einstein equation
\begin{align}
    R_{\mu \nu} = - \frac{6}{L_{\textrm{UV}}^2} g_{\mu \nu}.
\label{conditioncomparison2}
\end{align}
This leads to
\begin{align}
    \hat{f} &= - \frac{z^2}{20 \mathcal{\ell}^2} + \frac{31 z^4}{2000 \mathcal{\ell}^4} +\cdots,\nn \\
    \hat{g} &= \frac{z^2}{5 \mathcal{\ell}^2} - \frac{47 z^4}{1000 \mathcal{\ell}^4} + \cdots,
\label{fhatandghatcomparison2}
\end{align}
where $\mathcal{\ell}$ is the radius of $\mathbb{H}_3$, from which we extract
\begin{equation}
    \tilde f_1=\fft1{4\ell^2},\qquad \tilde f_2=-\fft{11}{200\ell^4}.
\end{equation}
The holographic entanglement entropy then takes the form
\begin{align}
    S^{\textrm{EE}}_{\textrm{UV}_{z \rightarrow 0}} \left( R; B^2 \times \mathbb{H}_3, \epsilon \right)
    &=\mathrm{Area}(\Sigma)
    \frac{L_{\textrm{UV}}^5}{4 G^{(7)}_N}
    \Bigg[\frac{1}{4\epsilon^4} + \left( \frac{1}{4 \mathcal{\ell}^2}- \frac{3}{64 R^2}  \right) \frac{1}{\epsilon^2}\nn \\
    &\kern3cm +\left( \frac{3}{200 \mathcal{\ell}^4} - \frac{3}{64 \mathcal{\ell}^2 R^2}-\frac{9}{2048 R^4}  \right) \log\left( \frac{R}{\epsilon} \right) \Bigg].
\label{holographicentanglemententropyAdS7toAdS4UVrofzcomparison2}
\end{align}
We now see that the $3/200\ell^4$ term in the log divergence indeed matches the CFT result, (\ref{termKappa0reduced}).  Moreover, we can identify the extrinsic curvature dependent coefficients in (\ref{termKappa0reduced}) as $\alpha_1=-3/64$ and $\alpha_2=-9/2048$.  Note, however, that there is as yet no direct field theory computation of these coefficients.

Finally, for the flowing theory from AdS$_7$ to AdS$_4\times\mathbb H_3$, the asymptotic form of the metric in the UV is given in (\ref{metricfunctionsorderfourAdS7toAdS4UV}).  This allows us to read off the coefficients
\begin{equation}
    \tilde f_1=\fft1{4\ell^2},\qquad\tilde f_2=-\fft{11}{200\ell^4}+\fft9{800\ell^4}=-\fft7{160\ell^4}.
\end{equation}
Note, in particular, that $\tilde f_1$ is unmodified, while $\tilde f_2$ picks up a additional factor of $9/800\ell^4$ from the relevant deformation.  Substituting these values into (\ref{holographicentanglemententropyAdS7toAdS4UVrofzcomparison1}) then yields perfect agreement with (\ref{eq:logcoef}). It would be quite interesting to augment the analysis of this section with the approach of \cite{Dong:2013qoa} which considered modifications to the entanglement entropy in presence of higher-derivative terms on the gravity side.

%%%%%%%%%%%%%%%%%%%%%%%%%%%%%%%%%%%%%%%%%%%%%%%%%%%%%%%%%
%%%%%%%%%%%%%%%%%%%%%%%%%%%%%%%%%%%%%%%%%%%%%%%%%%%%%%%%%%%%%%%%%%%%%%%%%%%%%%%%%%%%%%%%%%%%%%%%%%%%%%%%%%%%%%%%%%%%%%%%%%%%%%%%%%%%%%%%%%%%
\section{Conclusions}
\label{conclusions}
In this manuscript we continue explorations of holographic RG flows across dimensions. The analysis presented here is complementary to 
\cite{GonzalezLezcano:2022mcd,Deddo:2022wxj,Deddo:2023pid} where the focus was on holographic RG flows across dimensions triggered by magnetic fluxes along the compact manifold. Although it is clear that there is no na\"\i ve $c$-function across dimensions, we have pursued other notions with emphasis on preserving a subset of properties of the standard $c$-functions in flows in the same dimension. More precisely, here we have systematically studied
a holographic RG flow from $\textrm{AdS}_7$ to $\textrm{AdS}_4 \times \mathbb{H}_3$ and constructed versions of holographic entanglement entropies associated with this flow. 

We started in Section \ref{HolographicEE} by highlighting some of the key properties of RG flows across dimensions. In particular, these RG flows are holographically supported by magnetic fluxes and/or curvatures that cannot
be continuously turned off. Consequently, the field-theoretic
paradigm of deforming by a relevant deformation that can be turned off does not apply to these RG flows. Holographically, the discontinuity of these flows is clearly seen at the level of the equation of motion. This manuscript, together with 
\cite{GonzalezLezcano:2022mcd,Deddo:2022wxj,Deddo:2023pid}, clarifies that deformation by fluxes or deformation by curvature have  a similar role of triggering a Lorentz-violating RG flow whose deformation cannot be turned off continuously. 

In section~\ref{HolographicEE7to4} the construction of the solutions of the BPS equations was performed: sections \ref{BPSUV} and \ref{BPSIR} were devoted to the analytical solutions in the UV and IR regimes, respectively, while in section \ref{BPSnumerical} the numerical solutions were provided. The holographic entanglement entropy was constructed in detail in section \ref{HEE}: the construction associated with the UV regime was performed in section \ref{UVentanglemententropy} while the one associated with the IR regime can be found in section \ref{IRentanglemententropy}. In order to analyze the monotonicity properties of the holographic RG flows from $\textrm{AdS}_7$ to $\textrm{AdS}_4 \times \mathbb{H}_3$ it was necessary to construct the regulated holographic entanglement entropy in the whole spectrum as we move from the UV regime to the IR regime. For this purpose, a {\it Mathematica} code was written for two sets of data as described in section \ref{UVentanglemententropy-c} to capture the behavior of the regulated holographic entanglement entropy both in the UV and transition regimes, while another {\it Mathematica} code was written for a third set of data as described in section \ref{wholeentanglemententropy} to capture the behavior of the regulated holographic entanglement entropy in the IR regime. The behavior of the regulated holographic entanglement entropy in the whole spectrum is depicted in figure \ref{SEEreg}. Furthermore, the construction of a $c$-function associated with holographic RG flows from $\textrm{AdS}_7$ to $\textrm{AdS}_4 \times \mathbb{H}_3$ required the regulated holographic entanglement entropy constructed in Section \ref{Thecfunction} which is shown in Figure \ref{SEEreg2}. This $c$-function is monotonically decreasing along the holographic RG flow, it approaches the IR central charge in the IR regime and it blows up in the UV regime. This monotonically decreasing non-interpolating $c$-function is depicted in Figure \ref{cfunction2}. 

We also provided a field-theoretic interpretation of various terms appearing in the expression for the entanglement entropy by considering the coefficient of the logarithmic term in the context of a field theory placed in a curved spacetime and entangling regions determined by a generic region $\Sigma$. Namely, the new type of RG flows that we are tackling using holography are characterized by the field theory spacetime manifold on which the theory lives, $M$, and the geometry of the entangling region, $\Sigma$. We also want to accommodate the ubiquitous presence of fluxes along the compactification manifold that can be easily treated in the holographic setup. As discussed in section \ref{sec:anomalycoefs}, for a CFT defined on a 6d manifold $M$, the entanglement entropy is $S_{EE}(M,\Sigma, \epsilon)$; the coefficient of the logarithmic term is universal and depends on the Weyl tensor, the induced metric on $\Sigma$ and the intrinsic and extrinsic curvatures of $\Sigma$. The complete   schematic expression is given in Equation  (\ref{eq:a3schem}) which could also serve as a proxy for the central charged in the non-Lorentz-invariant theory.  To the best of our knowledge the full field-theoretic expression is not known in the literature. We provided a holographic treatment which led to the explanation of certain terms; the full field-theoretic answer remains an interesting open question.

%%%%%%%%%%%%%%%%%%%%%%%%%%%%%%%%%%%%%%%%%%%%%%%%%%%%%%%%%%%%%%%%%%%%%%%%%%%%%%%%%%%%%%%%%%%%%%%%%%%%%%%%%%%%%%%%%%%%%%%%%%%%%%%%%%%%%%%%%%%%%%%%%%%%%%%%%%%%%
\vskip 0.5 truecm
\section*{Acknowledgements}
We are grateful to Evan Deddo, Ioannis Papadimitriou and Christoph Uhlemann for various comments. This work is partially supported by the U.S. Department of Energy under grant DE-SC0007859. The work of J.~de-la-Cruz-Moreno is mostly supported by the Mexican Government through Secretar\'ia de Educaci\'on, Ciencia, Tecnolog\'ia e Innovaci\'on de la Ciudad de M\'exico (SECTEI). Partial funding was also provided by the Leinweber Center for Theoretical Physics (LCTP). LPZ gratefully acknowledges support from an IBM Einstein fellowship while at the Institute for Advanced Study.

%%%%%%%%%%%%%%%%%%%%%%%%%%%%%%%%%%%%%%%%%%%%%%%%%%%%%%%%%%%%%%%%%%%%%%%%%%%%%%%%%%%%%%%%%%%%%%%%%%%%%%%%%%%%%%%%%%%%%%%%%%%%%%%%%%%%%%%%%%%%%%%%%%%%%%%%%%%%%
\begin{appendices}

%%%%%%%%%%%%%%%%%%%%%%%%%%%%%%%%%%%%%%%%%%%%%%%%%%%%%%%%%%%%%%%%%%%%%%%%%%%%%%%%%%%%%%%%%%%%%%%%%%%%%%%%%%%%%%%%%%%%%%%%%%%%%%%%%%%%%%%%%%%%%%%%%%%%%%%%%%%%%
\section{Model associated with \texorpdfstring{$\kappa = 0$}{}}

In this section, to highlight the role of curvature, $\kappa$, we consider the case of $\kappa=0$; this way we are able to read off which contributions are universal an independent of the curvature.  We focus on clarifying which part of the result of reproducing the {\it Mathematica} code for the deep UV regime described in section \ref{UVentanglemententropy-c} is obtained from the contribution attributed to the $\kappa = 0$. Accordingly, we consider only the approximation where     \begin{align*}
    f_{\textrm{UV}}(z) &\approx -\log \left( z/L_{\textrm{UV}} \right), \\
    g_{\textrm{UV}}(z) &\approx -\log \left( z/L_{\textrm{UV}} \right).
    \numberthis
    \label{kappa0fg}
    \end{align*}

Reproducing all the procedure described in Section \ref{UVentanglemententropy} for the expression (\ref{kappa0fg}) we obtain that the leading order divergences of the holographic entanglement entropy are given by
    \begin{align*}
    S^{\textrm{EE}}_{\textrm{UV}_{z \rightarrow 0}} \left( R; B^2 \times \mathbb{H}_3, \epsilon \right)
    & =
    \frac{2 \pi \mathcal{\ell}^3 \textrm{vol}[\mathbb{H}_3]}{4 G^{(7)}_N}
    \Bigg[ 
    \displaystyle \int_\epsilon^{z_0} dz 
    \ r_{\textrm{UV}}(z) \left[ e^{2f_{\textrm{UV}}(z) + 3g_{\textrm{UV}}(z)} \right] \sqrt{1+(r_{\textrm{UV}}'(z))^2}
    \Bigg] \\ \\
    %%%%%%%%%%%%%%%%%%%%%%%%%%%%%%
    & =
    \frac{2 \pi \mathcal{\ell}^3 \textrm{vol}[\mathbb{H}_3]}{4 G^{(7)}_N}
    \Bigg[
    - \left( L_{\textrm{UV}}^5 R \right) \frac{1}{4 z^4}
    + \left( \frac{3 L_{\textrm{UV}}^5}{64 R} \right) \frac{1}{z^2} \\
    & \hspace{0.4cm}
    - \left( \frac{9 L_{\textrm{UV}}^5}{2048 R^3} \right) \log(z)
    + \mathcal{O} \left( z^2 \right)
    \Bigg] \Bigg|_{\epsilon}^{z_0} \\ \\
    %%%%%%%%%%%%%%%%%%%%%%%%%%%%%%
    & =
    \frac{2 \pi \mathcal{\ell}^3 \textrm{vol}[\mathbb{H}_3]}{4 G^{(7)}_N R^3}
    \Bigg[ L_{\textrm{UV}}^5
    \left[ \frac{R^4}{4 \epsilon^4} - \frac{R^4}{4 z_0^4} \right]
    - \frac{3}{64} L_{\textrm{UV}}^5
    \left[ \frac{R^2}{\epsilon^2} - \frac{R^2}{z_0^2} \right] \\
    & \hspace{0.4cm}
    - \frac{9}{2048}  L_{\textrm{UV}}^5
    \log \left( \frac{z_0}{\epsilon} \right)
    + \mathcal{O} \left( z^2 \right) \bigg|_{\epsilon}^{z_0} \Bigg]. \numberthis
    \label{holographicentanglemententropyAdS7toAdS4UVrofzkappa0}
    \end{align*}
This result helps interpret various terms in Equations  (\ref{holographicentanglemententropyAdS7toAdS4UVrofz}) and (\ref{holographicentanglemententropyAdS7toAdS4UVcoefficientsAB}). In particular, we notice precise numerical agreement with the $\kappa$-independent leading divergences in the coefficients $A$ and $B$ in (\ref{holographicentanglemententropyAdS7toAdS4UVcoefficientsAB}). The regulated holographic entanglement entropy in the deep UV regime has approximately the same behavior as expression (\ref{regulatedholographicentanglemententropyAdS7toAdS4UVnumerical}), 
    \begin{align}
    S^{\textrm{UV}}_{\textrm{EEreg}} \approx -0.34314 \frac{1}{R^3} -0.14062 \frac{1}{R^3} \log(R).
    \label{regulatedholographicentanglemententropyAdS7toAdS4UVnumericalkappa0}
    \end{align}

%%%%%%%%%%%%%%%%%%%%%%%%%%%%%%%%%%%%%%%%%%%%%%%%%%%%%%%%%%%%%%%%%%%%%%%%%%%%%%%%%%%%%%%%%%%%%%%%%%%%%%%%%%%%%%%%%%%%%%%%%%%%%%%%%%%%%%%%%%%%%%%%%%%%%%%%%%%%%
\section{Overlapping regions}
\label{consistency}
\renewcommand\thefigure{\thesection.\arabic{figure}}
\setcounter{figure}{0}

\begin{figure}[H]
\centering
    \begin{subfigure}[b]{0.47\textwidth}
    \centering
    \includegraphics[scale=0.8]{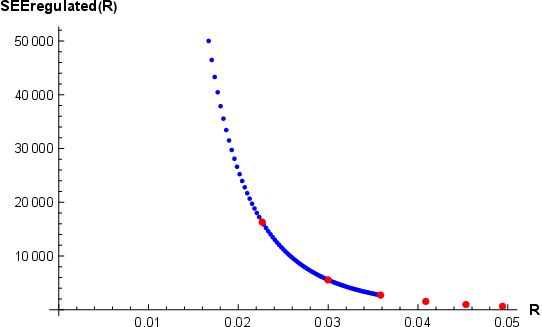}
    \caption{Matching of figures \ref{SEEreg-a} (blue dots) and \ref{SEEreg-b} (red dots) in the overlapping region of the UV regime and the transition regime.} %Both figures, \ref{SEEreg-a} and \ref{SEEreg-b}, were constructed with {\it Code 1}.}
    \label{ConsistencySEEreg-a}
    \end{subfigure}
    \hfill
    \begin{subfigure}[b]{0.47\textwidth}
    \centering
    \includegraphics[scale=0.8]{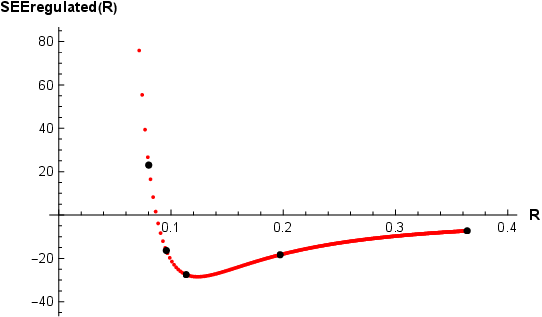}
    \caption{Matching of figures \ref{SEEreg-b} (red dots) and \ref{SEEreg-c} (black dots) in the overlapping region of the transition regime and the IR regime.} %Figures \ref{SEEreg-b} and \ref{SEEreg-c} were constructed with {\it Code 1} and {\it Code 2}, respectively.}
    \label{ConsistencySEEreg-b}
    \end{subfigure}
\caption{Matching of the regulated holographic entanglement entropy depicted in figure \ref{SEEreg} for both overlapping regions: a) UV and transition regimes, and b) transition and IR regimes.}
\label{ConsistencySEEreg}
\end{figure}

\begin{figure}[H]
\centering
    \begin{subfigure}[b]{0.47\textwidth}
    \centering
    \includegraphics[scale=0.8]{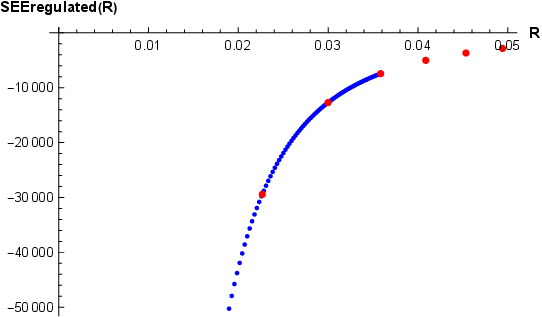}
    \caption{Matching of figures \ref{SEEreg2-a} (blue dots) and \ref{SEEreg2-b} (red dots) in the overlapping region of the UV regime and the transition regime.}
    \label{ConsistencySEEreg2-a}
    \end{subfigure}
    \hfill
    \begin{subfigure}[b]{0.47\textwidth}
    \centering
    \includegraphics[scale=0.8]{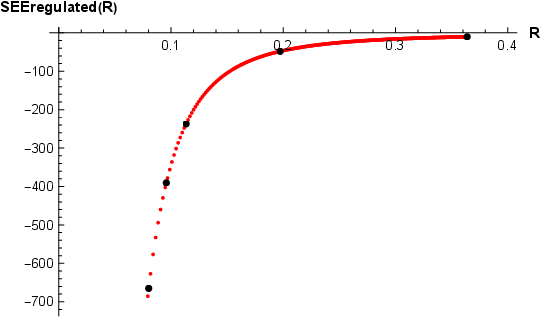}
    \caption{Matching of figures \ref{SEEreg2-b} (red dots) and \ref{SEEreg2-c} (black dots) in the overlapping region of the transition regime and the IR regime.}
    \label{ConsistencySEEreg2-b}
    \end{subfigure}
\caption{Matching of the regulated holographic entanglement entropy depicted in Figure \ref{SEEreg2} for both overlapping regions: a) UV and transition regimes, and b) transition and IR regimes.}
\label{ConsistencySEEreg2}
\end{figure}

\begin{figure}[H]
\centering
    \begin{subfigure}[b]{0.47\textwidth}
    \centering
    \includegraphics[scale=0.8]{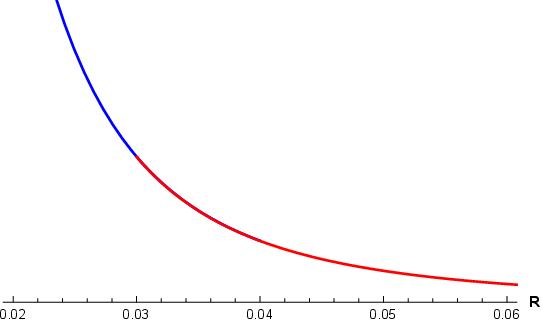}
    \caption{Matching of figures \ref{cfunction2-a} (blue line) and \ref{cfunction2-b} (red line) in the overlapping region of the UV regime and the transition regime.}
    \label{Consistencycfunction2-a}
    \end{subfigure}
    \hfill
    \begin{subfigure}[b]{0.47\textwidth}
    \centering
    \includegraphics[scale=0.8]{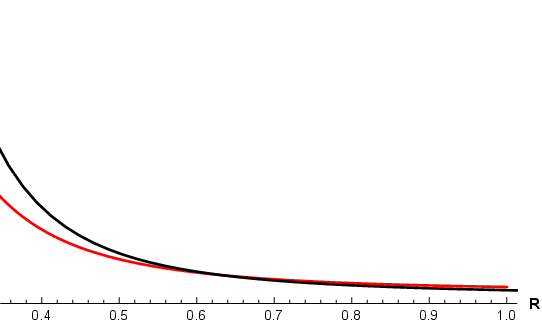}
    \caption{Matching of figures \ref{cfunction2-b} (red line) and \ref{cfunction2-c} (black line) in the overlapping region of the transition regime and the IR regime.}
    \label{Consistencycfunction2-b}
    \end{subfigure}
\caption{Matching of the $c$-function depicted in Figure \ref{cfunction2} for both overlapping regions: a) UV $c$-function and transition $c$-function, and b) transition $c$-function and IR $c$-function.}
\label{Consistencycfunction2}
\end{figure}

A crucial step in the constructions performed in this paper is the verification that the data match in the overlapping regions. Figures \ref{ConsistencySEEreg}, \ref{ConsistencySEEreg2}, and \ref{Consistencycfunction2}, stand for the verifications associated with Figures \ref{SEEreg}, \ref{SEEreg2}, and \ref{cfunction2}, respectively.

\end{appendices}

%%%%%%%%%%%%%%%%%%%%%%%%%%%%%%%%%%%%%%%%%%%%%%%%%%%%%%%%%%%%%%%%%%%%%%%%%%%%%%%%%%%%%%%%%%%%%%%%%%%%%%%%%%%%%%%%%%%%%%%%%%%%%%%%%%%%%%%%%%%%%%%%%%%%%%%%%%%%%
\vskip 0.5 truecm
\bibliographystyle{JHEP}
\bibliography{RG-Across-D}
\end{document}